\RequirePackage{fix-cm}
\documentclass[twocolumn,final]{svjour3}          
\usepackage[T1]{fontenc}
\usepackage[utf8]{inputenc}
\smartqed  

\usepackage{mathptmx}      
\usepackage{latexsym}

\usepackage{amsmath}
\usepackage{bm}

\usepackage{amsfonts}
\usepackage{amssymb}
\usepackage[usenames,pdftex,dvipsnames,svgnames,table]{xcolor}
\usepackage[final]{graphicx}

\usepackage{ulem}
\usepackage{hyperref}

\usepackage{xargs}                      
\usepackage[inline]{showlabels,rotating}

\usepackage[numbers,sort&compress]{natbib}

\usepackage{tikz}
\usetikzlibrary{hobby,patterns,arrows.meta}
\pgfdeclarepatternformonly{north east lines wide}%
   {\pgfqpoint{-1pt}{-1pt}}%
   {\pgfqpoint{6pt}{6pt}}%
   {\pgfqpoint{5pt}{5pt}}%
   {
     \pgfsetlinewidth{0.4pt}
     \pgfpathmoveto{\pgfqpoint{0pt}{0pt}}
     \pgfpathlineto{\pgfqpoint{5.1pt}{5.1pt}}
     \pgfusepath{stroke}
    }
\definecolor{applegreen}{rgb}{0.55, 0.71, 0.0}

\newcommand{\ub}{\mathbf{u}}

\newcommand{\xb}{\mathbf{x}}

\newcommand{\nb}{\mathbf{n}}

\newcommand{\be}[1]{
\begin{equation}
\expandafter\label{eq:#1}
}
\expandafter \newcommand{\ee}{\expandafter\end{equation}}
\expandafter \newcommand{\eq}[1]{Eq.~\expandafter(\ref{eq:#1})}

\expandafter \newcommand{\bfig}[1]{
\begin{figure}
\expandafter\label{fig:#1}
}
\expandafter \newcommand{\efig}{\end{equation}}
\expandafter \newcommand{\fig}[1]{Fig.~\ref{fig:#1}}

\newcommand{\diff}{{D}}

\renewcommand{\div}{\nabla\cdot}

\newcommand{\grad}{\nabla}
\newcommand{\gradn}{\nabla_{n}}

\newcommand{\spatavg}[1]{\left<#1\right>} 

\newcount\colveccount
\newcommand*\colvec[1]{
        \global\colveccount#1
        \begin{pmatrix}
        \colvecnext
}
\def\colvecnext#1{
        #1
        \global\advance\colveccount-1
        \ifnum\colveccount>0
                \\
                \expandafter\colvecnext
        \else
                \end{pmatrix}
        \fi
}

\newcount\rowveccount
\newcommand*\rowvec[1]{
        \global\rowveccount#1
        \begin{pmatrix}
        \rowvecnext
}
\def\rowvecnext#1{
        #1
        \global\advance\rowveccount-1
        \ifnum\rowveccount>0
                &
                \expandafter\rowvecnext
        \else
                \end{pmatrix}
        \fi
}

\renewcommand{\matrix}[1]{\begin{pmatrix}#1\end{pmatrix}}

\usepackage[obeyDraft,colorinlistoftodos,prependcaption,textsize=small]{todonotes}
\newcommandx{\mat}[2][1=]{\todo[inline,linecolor=red,backgroundcolor=blue!25,bordercolor=red,#1]{MATTEO: #2}}
\newcommandx{\ele}[2][1=]{\todo[inline,linecolor=blue,backgroundcolor=green!25,bordercolor=blue,#1]{ELEONORA: #2}}
\newcommandx{\gian}[2][1=]{\todo[inline,linecolor=OliveGreen,backgroundcolor=CarnationPink!25,bordercolor=OliveGreen,#1]{GIANLUCA: #2}}
\newcommandx{\alb}[2][1=]{\todo[inline,linecolor=purple,backgroundcolor=red!25,bordercolor=purple,#1]{ALBERTO: #2}}
\usepackage{soul}

\usepackage{xspace} 
\newcommand{\shm}{sHM\xspace}
\newcommand{\fhm}{fHM\xspace}
\newcommand{\shmU}{sHM\_U}
\newcommand{\shmR}{sHM\_R}
\newcommand{\fhmU}{fHM\_U}
\newcommand{\fhmR}{fHM\_R}

 \journalname{Computing and Visualization in Science}

\begin{document}

\title{Computational analysis of transport in three-dimensional heterogeneous materials
}
\subtitle{An \textsf{OpenFOAM}-based simulation framework}


\author{Gianluca Boccardo \and
        Eleonora Crevacore \and
        Alberto Passalacqua \and
        Matteo Icardi 
}


\institute{G. Boccardo \at
              Dipartimento di Scienze Applicate e Tecnologia,\\
              Politecnico di Torino, \\
              10129, Torino, Italy \\
              \email{gianluca.boccardo@polito.it}           
           \and
         	  E. Crevacore \at
              Dipartimento di Scienze Matematiche "G. L. Lagrange", \\
              Politecnico di Torino,\\
              10129, Torino, Italy\\
              \email{eleonora.crevacore@polito.it}
           \and
              \textbf{M. Icardi} (corresponding author) \at
         	  School of Mathematical Sciences, \\
         	  University of Nottingham, \\
         	  NG8 3LS, Nottingham, UK \\
         	  \email{ matteo.icardi@nottingham.ac.uk}
           \and
              A. Passalacqua \at
         	  Department of Mechanical Engineering, \\
         	  Iowa State University, \\
         	  Ames, IA 50011, USA \\
         	  \email{albertop@iastate.edu}
}

\date{Received: 2 June 2017 / Accepted: date}

\maketitle

\begin{abstract}
Porous and heterogeneous materials are found in many applications from composites, membranes, chemical reactors, and other engineered materials to biological matter and natural subsurface structures. In this work we propose an integrated approach to generate, study and upscale transport equations in random and periodic porous structures. The geometry generation is based on random algorithms or ballistic deposition. In particular, a new algorithm is proposed to generate random packings of ellipsoids with random orientation and tunable porosity and connectivity. The porous structure is then meshed using locally refined Cartesian-based or unstructured strategies.
Transport equations are thus solved in a finite-volume formulation with quasi-periodic boundary conditions to simplify the upscaling problem by solving simple closure problems consistent with the classical theory of homogenisation for linear advection-diffusion-reaction operators.
Existing simulation codes are extended with novel developments and integrated to produce a fully open-source simulation pipeline. A showcase of a few interesting three-dimensional applications of these computational approaches is then presented. Firstly, convergence properties and the transport and dispersion properties of a periodic arrangement of spheres are studied. Then, heat transfer problems are considered in a pipe with layers of deposited particles of different heights, and in heterogeneous anisotropic materials. 

\end{abstract}


\section{Introduction}
\label{intro}
In recent years computer simulations are becoming an ever more important tool to study a large spectrum of physical problems, a trend supported by the  increasing processor throughput and advances in the parallel computing research.
Nonetheless, while the constraint due to the cost in computational time is gradually easing, other bottlenecks can remain, such as the limited or expensive access to licensed (i.e. commercial) or otherwise closed-source codes, that are hardly verifiable and rigorously validated. While several promising open-source libraries and codes (Deal.II, UG, dune, GetFEM, just to cite a few) are now available,  there is still an urgent need of developing specific tools and workflow to approach particular classes of problems.
This is particularly problematic in the case of very complex physics or inherently multi-scale problems such as heterogeneous media where, despite the current growing trend in non-physical data-driven science, computer modelling still remains the only practical path to an improved knowledge and understanding and, eventually, design and optimisation. These problems, in fact, often require, not only efficient discretisation and solvers, but also specific features, and a wide range of pre- and post-processing tools.
Porous media, in their wide interpretation, are a prominent category of multi-scale problems, and, as such, form the basis and the motivation of the model problems explored in this paper~\cite{joekar2016pore,perovic2017multi}.

The purpose of this work is thus to present a series of tools and methods which can be used to approach a wide range of problems and applications in porous media, characterised by different geometrical descriptions and involved in several transport phenomena and dynamics: from transport of dilute suspensions to heat transfer in porous structures.
Eventually, the aim of this paper is to propose a fully open-source simulation workflow, show the validity and feasibility of a fully \textit{in-silico} investigation, and introducing some specific novel methods and formulation to streamline the modelling and upscaling step.
We demonstrate this approach by solving three proof-of-concept problems.
Although these are sometimes characterised by some ideal assumptions (e.g., geometrically simplified models, mostly linear and constant parameters, decoupled multi-physics), the tools presented in this work can be readily used for a number of important applications.
\mat{Potremmo citare un po' di applicazioni: batterie, reattori catalitici, filtri, fluidised beds, CO2 storage, membrane?}

Within the computational workflow mentioned above, we thus present here specific developments and advancements to:
\begin{itemize} 
\item generate realistic packings of solid grains of arbitrary shapes, suitable for the modelling of both natural formations (e.g. aquifers) and artificial packings (e.g. packed bed reactors), a class of unit operations central to process engineering~\citep{P003};

\item generate random arrangements of spheres or ellipsoids with a tunable porosity, which can be used  to used to represent realistic representations of solid dispersions with a variable amount of dispersed matter (as it will be shown in this work);

\item improve on the classic boundary conditions used in unsteady linear transport problems, to strongly reduce computational times in the investigation of the upscaled dynamics of a quasi-periodic problem, bringing a definite improvement on the ``naive'' methods, especially for problems characterised by very large scales of variation of the quantity of interest (e.g.: advection-diffusion problems at very high P\'eclet numbers); this technique was already successfully used in a recent work of ours~\citep{boccardo2018robust};

\item define and solve efficiently closure cell problems for diffusion in heterogeneous anisotropic materials.

\end{itemize}

Other mathematical models and open-source simulation tools, based on the same computational platform, can be also profitably employed in conjunction to the methods explored here, to overcome some of the limitations.
These include a quadrature-based method of moments to deal with the solution of PDF transport equations \citep{openqbmm} (e.g., evolution of particle populations or other additional internal coordinates such as transit time), and stochastic sampling techniques based on multi-level Monte Carlo methods to deal with uncertainty quantification and stochastic upscaling in heterogeneous porous media systems~\citep{icardi2016predictivity}.

The structure of the paper is thus the following: in Section \ref{sec:eq} the theoretical bases of the investigated problems are laid down, given that the real-world test cases explored share the same theoretical background, fully or in part.
Section~\ref{sec:numerics} summarises the numerical techniques, based on a finite-volume formulation.
Eventually, the actual computational workflow is detailed in Section \ref{sec:results} for each case: from the generation of the geometrical model, to the creation of a suitable mesh, to the presentation of the actual results.
Specifically, the first part will deal with transport of a solute in porous medium composed of a periodic arrangement of solid grains, with added attention to the meshing strategies employed in the simulation.
The second part explores fluid flow and heat transfer in a pipe where varying amounts of solid matter has settled on its bottom, focusing on the resulting effect on the heat transfer coefficient of the system.
Lastly, another heat transfer problem is presented, this time considering the case of a solid dispersion, where the continuous matrix and the dispersed inclusions are characterised by different thermal diffusivity.
In this way, the reader will be able to more effectively peruse this work, focusing on the application of interest and on the specific computational pipeline for its development.

\section{Model equations}
\label{sec:eq}

Let us consider a generic two-phase heterogeneous material (cf. Fig.~1), denoted with $\Omega=\Omega_{1}\cup\Omega_{2}$, $\Gamma$ being the interface between them, and let us split the external boundary in the inlet $\partial\Omega_{-}$, outlet $\partial\Omega_{+}$, and lateral boundary $\partial\Omega_{\ell}$.

\begin{figure}[h!]
\centering

\begin{tikzpicture}[use Hobby shortcut,closed=true,scale=1.5]

\draw[fill=gray!40!] (-1.45,1.4)--(1.35,1.4)--(1.35,-1.4)--(-1.45,-1.4);
\draw[black, line width=0.25mm]  (1.35,1.4)--(1.35,-1.4);
\draw[black, line width=0.25mm]  (1.38,1.4)--(1.38,-1.4);
\node at (1.7,0) {{\large $\partial \Omega+$}};

\draw[black, line width=0.25mm]  (-1.45,1.4)--(-1.45,-1.4);
\draw[black, line width=0.25mm]  (-1.48,1.4)--(-1.48,-1.4);
\node at (-1.8,0) {{\large $\partial \Omega-$}};

\draw[blue, line width=0.45mm]  (-1.45,1.4)--(1.35,1.4);
\node[blue] at (1,1.55) {{\large $\partial \Omega_{\ell}$}};
\draw[blue, line width=0.45mm]  (-1.45,-1.4)--(1.35,-1.4);

\draw[fill=white, postaction={pattern=north east lines wide}] (0,0).. (-0.2,0).. (-0.4,0.2) .. (-0.2,0.7).. (-0.1,0.6).. (0.2,0.5).. (0.3,0.4).. (0.3,0.2).. (0,0);
\draw[line width=0.5mm, red] (0,0).. (-0.2,0).. (-0.4,0.2) .. (-0.2,0.7).. (-0.1,0.6).. (0.2,0.5).. (0.3,0.4).. (0.3,0.2).. (0,0);

\draw[fill=white, postaction={pattern=north east lines wide}] (-0.2,-0.2).. (-0.2,-0.4).. (-0.3,-0.5) .. (-0.65,-0.5).. (-0.65,-0.2).. (-0.5,-0.05).. (-0.2,-0.2);
\draw[line width=0.5mm, red] (-0.2,-0.2).. (-0.2,-0.4).. (-0.3,-0.5) .. (-0.65,-0.5).. (-0.65,-0.2).. (-0.5,-0.05).. (-0.2,-0.2);

\draw[fill=white, postaction={pattern=north east lines wide}] (0.2,-0.25)..(0.4,-0.3)..(0.6,-0.5)..(0.4,-0.7)..(0,-0.5)..(0.2,-0.25);   
\draw[line width=0.5mm, red] (0.2,-0.25)..(0.4,-0.3)..(0.6,-0.5)..(0.4,-0.7)..(0,-0.5)..(0.2,-0.25);

\draw[fill=white, postaction={pattern=north east lines wide}] (0.7,-0.25)..(0.5,-0.1)..(0.7,0.35)..(1,0.3)..(1.2,0.1)..(0.7,-0.25);
\draw[line width=0.5mm, red] (0.7,-0.25)..(0.5,-0.1)..(0.7,0.35)..(1,0.3)..(1.2,0.1)..(0.7,-0.25);

\draw[fill=white, postaction={pattern=north east lines wide}] (0.55,0.6)..(0.5,0.8)..(0.9,1.2)..(1.1,0.75)..(0.8,0.5)..(0.55,0.6);
\draw[line width=0.5mm, red] (0.55,0.6)..(0.5,0.8)..(0.9,1.2)..(1.1,0.75)..(0.8,0.5)..(0.55,0.6);

\draw[fill=white, postaction={pattern=north east lines wide}] (-0.22,1.1)..(-0.2,0.85)..(-0.1,0.75)..(0.2,0.75)..(0.4,0.9)..(-0.22,1.1);
\draw[line width=0.5mm, red] (-0.22,1.1)..(-0.2,0.85)..(-0.1,0.75)..(0.2,0.75)..(0.4,0.9)..(-0.22,1.1);

\draw[fill=white, postaction={pattern=north east lines wide}] (1.3,-0.5)..(1.1,-0.4)..(1,-0.5)..(0.55,-0.9)..(0.5,-1)..(0.6,-1.1)..(1.3,-0.5);
\draw[line width=0.5mm, red] (1.3,-0.5)..(1.1,-0.4)..(1,-0.5)..(0.55,-0.9)..(0.5,-1)..(0.6,-1.1)..(1.3,-0.5);

\draw[fill=white, postaction={pattern=north east lines wide}] (0.2,-0.8)..(0.05,-0.7)..(-0.2,-0.65)..(-0.4,-0.7)..(0.2,-0.8);
\draw[line width=0.5mm, red]  (0.2,-0.8)..(0.05,-0.7)..(-0.2,-0.65)..(-0.4,-0.7)..(0.2,-0.8);

\draw[fill=white, postaction={pattern=north east lines wide}] (-0.8,-1.05)..(-0.7,-0.6)..(-1,-0.4)..(-1.3,-1)..(-0.8,-1.05);
\draw[line width=0.5mm, red] (-0.8,-1.05)..(-0.7,-0.6)..(-1,-0.4)..(-1.3,-1)..(-0.8,-1.05);

\draw[fill=white, postaction={pattern=north east lines wide}] (-1,0)..(-0.6,0.2)..(-0.6,0.3)..(-0.9,0.45)..(-1.2,0.4)..(-1.2,0.1)..(-1,0);
\draw[line width=0.5mm, red] (-1,0)..(-0.6,0.2)..(-0.6,0.3)..(-0.9,0.45)..(-1.2,0.4)..(-1.2,0.1)..(-1,0);

\draw[ red , line width=0.5mm, fill=white,postaction={pattern=north east lines wide}] (-1.3,0.8)..(-1,0.65)..(-0.9,0.7)..(-0.6,0.7)..(-0.6,1.1)..(-0.9,1.2)..(-1.3,0.8);

\node[red] at (0.35,0.6) {{\large $\Gamma$}};

\end{tikzpicture}
\label{fig:cartoon}
\caption{Schematic representation of a representative volume element of a generic two-phase heterogeneous material.}

\end{figure}

In the following, we will consider a few simplifications, the chief of which is to assume a saturated porous medium, constituted by solid non-connected irregular grains immersed in a continuous phase, either liquid (i.e. Newtonian incompressible fluid at room temperature) or solid (with different characteristics from the grains).

\subsection{Flow field}
The equations governing the fluid phase are the stationary Navier-Stokes equation\footnote{Body forces are neglected.} and the mass balance law:
\be{Navier-Stokes}
	\rho (\nabla \mathbf{u}) \mathbf{u} = -\nabla p + \mu \Delta \mathbf{u} \ , 
\ee
\be{mass-balance}
	\nabla \cdot \mathbf{u} = 0 \ ,
\ee
where $\rho$ is the constant fluid density (kg m$^{-3}$), $\mathbf{u}$ is the effective fluid velocity (m s$^{-1}$), $p$ is the pressure (kg m$^{-1}$ s$^{-2}$) and $\mu$ is the fluid dynamic viscosity (kg m$^{-1}$ s$^{-1}$).
When inertia is negligible, \eq{Navier-Stokes} reduces to the Stokes's equation:
\be{Stokes}
-\nabla p + \mu \Delta \mathbf{u} = 0\ , 
\ee
this case is usually referred to as ``creeping flow'', typically encountered in several environmental applications: \hl{all our simulations are performed in this range, in laminar conditions}.

\subsection{Transport models}
Coupled to the fluid phase, we generally solve a transport problem assuming it does not have a back-coupling with the flow. This is true for small dilute particles or for solutes. When no flow is present (i.e. the continuous phase is a solid), there is no need of solving the flow field, although in many applications advection can be due to electrostatic forces or other physics which we neglect here.

\subsubsection{Eulerian equations}
Working in an Eulerian framework, it is possible to model the concentration transport of solutes or of a diluted suspension of colloidal particles as:
\be{Trasporto}
\frac{\partial c}{\partial t} + \mathbf{u} \cdot \nabla c = \nabla \cdot (D \nabla c) \ ,
\ee
where $c$ is the concentration (kg m$^{-3}$), $D$ is the diffusion coefficient (m$^2 $s$^{-1}$); for the advective term we made use of the incompressibility condition \eq{mass-balance}.
As it has been said, the interactions between solid and fluid phases are described with \textit{one-way coupling}, stating that the fluid (and external forces) will affect the motion of the particles, but not the vice-versa. The diffusion coefficient for small particles can be estimated with the Stokes-Einstein equation:
\be{Stokes-Einstein}
D = \frac{\kappa_{\textrm{B}}T}{3 \pi \mu d_{\textrm{p}} }\ ,
\ee
where $\kappa_\textrm{B}$ is the Boltzmann constant, $T$ is the temperature (K) and $d_\textrm{p}$ is the particle diameter (m).
When dealing with solid phases (inside the grains or in both domains), $\mathbf{u}=0$ and $D$ is a solid diffusion coefficient (possibly a tensor).

It is worth noting that an equation of the form of \eq{Trasporto} can apply for heat transfer problems (studied in Sections \ref{sec:pipeSand} and \ref{sec:soliDisp}) as well.
In this case, it is convenient to introduce the following notation:
\begin{equation}\label{eq:heatTransfer}
\frac{\partial T}{\partial t} + \mathbf{u} \cdot \nabla T = \nabla \cdot (\alpha \nabla T) \ ,
\end{equation}
where T is the fluid temperature (K) and the diffusive coefficient is now the thermal diffusivity (scalar or tensor) $\alpha$ (m$^2 $s$^{-1}$).

\subsection{Scalar quasi-periodic boundary conditions}\label{sec:bc}
Eulerian transport problems, even in the case of periodic geometry and flow,
cannot be solved with simple periodic conditions on a single periodic unit cell due
 to intrinsic evolution in space and time of the equations that requires larger domain to be solved, or
 due to the non-conservative nature of the equations.
 However, when the equation is linear and we are solely interested in the asymptotic
 (long-time, infinitely far, self-similar) behaviour, we can reformulate the problem to
 find a stationary \textit{quasi-periodic solution} (up to a multiplicative or additive constant, depending on the nature of the problem). Dropping the time dependence, we can formulate a generic linear transport problem as
 \be{quasi-periodic}
 \div{(\mathfrak{L}c)}=\mathfrak{R}c+\mathfrak{F}\qquad \mbox{in}\quad\Omega
 \ee
 where $\mathfrak{L}$ is a flux operator (e.g., advection-diffusion fluxes), $\mathfrak{R}$ is a non-conservative operator (e.g., reactions, differential operators not in a divergence form), $\mathfrak{F}$ is a source term. These operators are all assumed to be linear in $c$ but possibly dependent on space $x$.
We consider the following internal boundary conditions\footnote{On the porous matrix walls, in the case of perforated domain. Otherwise, for dual domain problems, this has to be replaced with interface conditions and an additional equation for the other domain has to be considered.}:
 \be{quasi-periodic-bc}
 \mathfrak{L} c = \mathfrak{f}(c)\qquad \mbox{on}\quad\Gamma
 \ee
with $\mathfrak f$ being a generic (space-dependent, possibly non-linear) flux at the wall.

We aim to find a self-similar solution of \eq{quasi-periodic}, i.e., a solution in the smallest periodic cell with quasi-periodic external boundary conditions of the type\footnote{Depending on the form of the flux operator, the resulting PDE can be elliptic, parabolic or hyperbolic. We consider here conditions both for the solution and its normal gradient, generally needed for elliptic PDEs.}:
 \be{quasi-periodic-bc2}
 c|_{\partial\Omega^-} =  \phi(c|_{\partial\Omega^+}), \qquad
 \gradn c|_{\partial\Omega^-} =  \psi(c|_{\partial\Omega^+})
 \ee
 where $\phi$ and $\psi$ are linear functions and $\partial\Omega^\pm$ are two (geometrically opposite) periodic boundaries\footnote{In three dimensional cubical elementary volumes, one can write separate equations for $x,y,z$ periodicity.}. It follows that, generally, $\psi(c)=\phi'(c)\gradn c$ to ensure the quasi-periodicity of the normal gradient. From simple compatibility conditions for the existence and uniqueness of solutions, we can distinguish the following cases:
 \begin{itemize}
 \item \textit{Conservative transport:} $\mathfrak{R}=\mathfrak{F}=\mathfrak{f}=0$. This setup is the one of interest for studying \textit{passive asymptotic dispersion} (or effective diffusion when there is no advection) properties as what remains is a fully conservative equation. The quantity of interest to quantify the dispersion is the average of local gradients $\grad c$. However, for periodic conditions, only a trivial constant solution (with $\phi(c)=\psi(c)=c$) exists. Therefore, we have to re-introduce a fictitious non-conservative constant source $\mathfrak{f}=\mathfrak{f}_{0}=\mathfrak{L}(\mathbf{p}\cdot\xb)$ where  $\mathbf{p}$ is the direction in which we want to study the transport ($x$ in this case). This source term is zero for pure diffusion. The resulting equation has now a non-trivial solution up to an additive constant. We therefore
 select a quasi-periodicity of the type:
 \be{cons-bc}
 \phi(c)=c+\phi_{0}, \qquad \phi_{0}=\spatavg{c}_{\partial\Omega^+}-\spatavg{c}_{\partial\Omega^-}
 \ee
 where $\spatavg{\cdot}$ represents an averaging (surface- or flux-weighted) operator\footnote{In fact, considering a linear and periodic operator $\mathfrak{L}$, surface and flux-weighted averages are equivalent, i.e., $\spatavg{\mathfrak{L}c}=\spatavg{c}$.}, and the constant $\phi_{0}$
is a variable of the problem and it has to be computed to counter-balance the volumetric source term. Without loss of generality we can assume then $\spatavg{\phi(c|_{\partial\Omega^-})}=0$. We can then compute the local gradients (whose norm gives the dispersion coefficient), produced by the transport operator $\mathfrak{L}$ under the macroscopic gradient $\phi_{0}$, by subtracting the source term $\mathfrak{f}_{0}$ as
 $$
\frac{ \grad c - \grad{(\mathbf{p}\cdot \xb)}}{\phi_{0}}
 $$
 \item \textit{Homogeneous equation:} $\mathfrak{F}=0$ and $\mathfrak{f}(c)=\mathfrak{f}_{1}c$. This is the case of linear bulk and surface reaction, where, in general, the solution is given up to a multiplicative constant. This suggests us to look for quasi-periodic solutions with 
 \be{hom-bc}
 \phi(c)={\phi_{1}}{c}, \qquad
 \phi_{1} = \frac{\spatavg{\mathfrak{L}c}_{\partial\Omega^-}}{\spatavg{\mathfrak{L}c}_{\partial\Omega^+}}
 		=\frac{\spatavg{c}_{\partial\Omega^-}}{\spatavg{c}_{\partial\Omega^+}}
 \ee
in such a way that the quasi-periodic boundary conditions balance the non-conservative terms and allows to identify $\phi_{1}$ as an equivalent reaction rate (observing it can be related to the volume-averaging of all non-conservative terms).
 
  \item \textit{General linear case:}
  Integrating \eq{quasi-periodic} over the volume, and considering a standard advection diffusion operator
  $\mathfrak{L}=\ub+\diff\grad$, linear bulk reaction $\mathfrak{R}$, linear surface reaction
  $$
  \mathfrak{f}(c)=\mathfrak{f}_{1}c+\mathfrak{f}_{0}
  $$
  and a linear transformation $\phi(c)=\phi_{1}c+\phi_{0}$, the following conditions hold for $\phi_{1}$
  \begin{eqnarray}
  \nonumber
  \phi_{1} &=& \frac{\spatavg{\mathfrak{L}c}_{\partial\Omega^-}}{\spatavg{\mathfrak{L}c}_{\partial\Omega^+}} = 1 +
  \mathfrak{R}\frac{\int_{\Omega}{c}}{\int_{\partial\Omega^+}{\mathfrak{L}c}}+
  \mathfrak{f_{1}}\frac{\int_{\Gamma}{c}}{\int_{\partial\Omega^+}{\mathfrak{L}c}}\\
  \label{eq:gen-bc}
  &=& 1 +
  \mathfrak{R}\frac{\spatavg{c}}{\spatavg{\mathfrak{L}c}_{\partial\Omega^+}}\frac{|\Omega|}{|\partial\Omega^+|}+
  \mathfrak{f_{1}}\frac{\spatavg{c}_\Gamma}{\spatavg{\mathfrak{L}c}_{\partial\Omega^+}}\frac{|\Gamma|}{|\partial\Omega^+|}
  \end{eqnarray}
and $\phi_{0}$
  \begin{eqnarray}
  \nonumber
  \phi_{0} &=&
  \mathfrak{F}\frac{\int_{\Omega}1}{\int_{\partial\Omega^-}\mathfrak{L}1}+
  \mathfrak{f_{0}}\frac{\int_{\Gamma}1}{\int_{\partial\Omega^-}\mathfrak{L}1}\\
  \label{eq:gen-bc2}
  &=&
  \mathfrak{F}\frac{1}{\spatavg{\ub\cdot\nb}_{\partial\Omega^-}}\frac{|\Omega|}{|\partial\Omega^-|}+
  \mathfrak{f_{0}}\frac{1}{\spatavg{\ub\cdot\nb}_{\partial\Omega^-}}\frac{|\Gamma|}{|\partial\Omega^-|}
  \end{eqnarray}
As it can be seen from the last equality, the integrals can be rewritten highlighting the geometric factors $\frac{|\Omega|}{|\partial\Omega^-|}$ and $\frac{|\Gamma|}{|\partial\Omega^-|}$, and the mean concentration values $\spatavg{c}$ in the volume and on the surfaces. If the flow field is incompressible, $\spatavg{\ub\cdot\nb}_{\partial\Omega^-}$ is equivalent to the mean (volumetric) velocity.
  \item \textit{Non-linear case:}
  For general non-linear operators, one has to find a function $\phi(c)$ such that
  $$
  {\int_{\partial\Omega^+}{\mathfrak{L}\phi(c)}} = \int_{\partial\Omega^+}{\mathfrak{L}c} +
  {\int_{\Omega}{\mathfrak{R}c}}+
  {\int_{\Gamma}{\mathfrak{f}(c)}}
  $$
  This problem is not generally easily solvable. One possibility is to assume a ``macroscopically'' linear reaction, i.e., $\phi$ to be linear. If $\phi(c)=\phi_{1}c$, the problem of finding $\phi_{1}$ becomes equivalent to a generalised non-linear eigenvalue problem. There is no guarantee, however, that the problem admits a solution. More in general, in our future works, we will investigate a more general algorithm to find linear and non-linear functions $\phi$.
 \end{itemize}

This  proposed approach can be proven to be equivalent to the several cell problems formally derived by two-scale expansions \citep{hornung2012homogenization}, and it is particularly convenient in the cases where the phenomenon of interest would require very long physical time or especially, a very large spatial domain. This is true in most upscaling problems where, to derive upscaled equations, the asymptotic regime is often sought. Simulations on a smaller periodic domain with these sets of \textit{quasi-periodic boundary conditions} can significantly reduce the computational effort.
It is interesting to notice that, in two-scale asymptotic homogenisation, this is usually overcome with the formal derivation of cell problems with periodic conditions. Our approach, despite being more phenomenological, is indeed equivalent and can be applied also to problems where an homogenisation limit does not exist.
From the computational point of view, \eq{cons-bc}, \eq{hom-bc}, and \eq{gen-bc}, have to be implemented via outer iterations, with $\phi_{1}$ and $\phi_{0}$ computed from the field $c$. In our implementation, outer iterations are always present also to allow for non-linear terms, non-linear discretisation schemes, and explicit corrections for non-orthogonal cells. At each iteration, the boundary conditions are computed with the equations above. The iterations stop when the residuals of the equations fall below a certain threshold and, at the same time, the estimated pseudo-periodic boundary conditions converges to a constant value.

\section{Numerical discretisation}\label{sec:numerics}
%

\subsection{Numerical schemes}
\textsf{OpenFOAM} implements the finite volume method (FVM) \cite{FerzigerPeric2002} with co-located grid arrangement.
Internal values are stored at the cell centre, while boundary values are stored at the face centroid on the corresponding boundary cell faces.
The Rhie-Chow approach \cite{RhieChow1983,shenImproved2001} is used to address the pressure-velocity decoupling observed when this grid arrangement is adopted.
When computing fluxes, values of each variable are computed at the face centroid of each interior cell through a reconstruction technique, following \cite{darwish_tvd_2003} (more details on the implementation can be found in \cite{jasak1996error}), as shown below.

The implementation of numerical schemes for convection and diffusion schemes follows the standard FVM approach \cite{FerzigerPeric2002,Darwish2016}. 

The convective term is discretised by applying Gauss' theorem. If we consider a scalar $c$, advected with a velocity $\mathbf{u}$, and we indicate the volume of a computational cell with $V$, while $\partial V$ is its surface area, we can write
\begin{equation}
\int_V \nabla \cdot \left( \mathbf{u} c \right) \textrm{d}V = \int_{\partial V} (\mathbf{u} c) \cdot \mathbf{n}\  \textrm{d}S \approx \sum_f c_f \mathbf{u}_f \cdot \mathbf{n}_f S_f.
\end{equation}
%
The value of both $\mathbf{u}_f$ and of $c_f$ need to be evaluated at cell faces; $S_f$ is the surface area of the considered faces and $\mathbf{n}_f$ is the associated outward unit vector. If a second-order TVD scheme is used, this is achieved by means of the reconstruction procedure illustrated in \cite{darwish_tvd_2003} (Sec. 2), in which the face value of a variable is obtained as
\begin{equation}
c_i = c_P + \frac{1}{2}\psi(r_f) (c_W - c_P),
\end{equation}
where $c_P$ is the value of $c$ at the cell centre (labelled $P$, as illustrated in Fig.~\ref{fig:cartoon}), $c_W$ the one at the downwind node, and $\psi(r_f)$ is the limiter function. The quantity $r_f$ is defined as \cite{darwish_tvd_2003}
\begin{equation}
r_f = \frac{c_P - c_E}{c_W - c_P},
\end{equation}
where $c_E$ is the upwind value of $c$.

The advantage of this procedure consists in its generality, which allows the first-order upwind scheme and the central differencing scheme to be recovered by setting $\psi(r_f) = 0$ and $\psi(r_f) = 1$ respectively, \cite{darwish_tvd_2003}.

The diffusion term is discretised considering the Laplacian operator \cite{OpenFoamProgrammersGuide}
\begin{equation}
\int_V \nabla \cdot \left( \Gamma \nabla c \right) \ \textrm{d} V = \int_{\partial V} \left( \Gamma \nabla c \right) \cdot \mathbf{n}\ \textrm{d}S \approx \sum_f \Gamma_f \left(\nabla c \right)_f \cdot \mathbf{n}_f S_f,
\end{equation}
with
\begin{equation}
\left(\nabla c \right)_f \cdot \mathbf{n}_f = \frac{c_E - c_P}{|\mathbf{d}|} S_f \ \cos \theta + \mathcal{C}_\textrm{g},
\label{eq:FaceNGrad}
\end{equation}
where 
$\mathbf{d}$ is the vector joining the cell centre of the considered adjacent cells.
The term $\mathcal{C}_\textrm{g}$ is an explicit correction for non-orthogonality \cite{Darwish2016}.

Gradients are computed either with Gauss' integration
\begin{equation}
\int_V \nabla c \ \textrm{d} V \approx \sum_f c_f S_f, 
\label{eq:GaussGrad}
\end{equation}
where the face values of $c$ are found using a linear reconstruction of the cell values on cell faces, or using the least-squares approach. Several implementations of the latter are available in \textsf{OpenFOAM}: the second-order least-squares approach, available in different flavours, either using the cell-centred values or the centre and nodal values for the calculation, and a fourth-order least-square approach.

The adoption of Gauss' gradient approach (Eq.~\eqref{eq:GaussGrad}) is only appropriate on regular meshes, where it has second-order accuracy (see, among others~\cite{FerzigerPeric2002}).
Additionally, this approach is sensitive to mesh anisotropy. Consequently, on meshes with varying cell size, mesh refinement and arbitrarily shaped cells, the least-squares approach is the recommended choice to ensure second-order accuracy. Thus, we adopt it when dealing with irregular meshes, as described in the following cases.

Another source of potential difficulties is the presence of the explicit non-orthogonal correction in the calculation of the face-normal gradient (Eq.~\eqref{eq:FaceNGrad}), which may lead to unbounded results, depending on the mesh quality and on the problem being solved; the difficulty with the correction term is that, being it explicit, it may lead to unboundedness when the non-orthogonality is too  large, with consequent loss of robustness of the scheme.
The approach used in OpenFOAM is presented in~\cite{jasak1996error}.

To avoid unboundedness, one can either apply of the correction only when mesh non-orthogonality is high, or exclude the correction altogether.
Direct consequence of this is the reduction of the accuracy of the numerical scheme, which deteriorates when the non-orthogonality problem is particularly serious.


\begin{figure}[h!]
\centering
\begin{tikzpicture}[use Hobby shortcut,closed=true,scale=0.45]

\draw[black, -{Stealth[length=2mm]}] (2,8) -- (6,8);
\node[black] at (4,7.3) {{\large Wind direction}};

\draw[black] (0,5) -- (0,0) -- (5,0) ;
\draw[black] (6,6.5) -- (1.75,6.2) -- (0,5) -- (3,5) ;
\draw[black, dashed] (1.75,6.2)--(2,2)--(7.5,1.5);
\draw[black, dashed] (2,2)--(0,0);
\draw[fill=gray!20!, fill opacity=0.2] (5,0)--(7.5,1.5)--(6,6.5)--(3,5)--(5,0);
\draw[black, line width=0.45mm] (2.85,3)--(3.15,3);
\draw[black, line width=0.45mm] (3,2.85)--(3,3.15);
\node at (2.5,3) {{\large $P$}};

\node at (1.65,0.5) {{\large $V$}};
\draw[-{Stealth[length=2mm]}] (5.66,3.4) -- (7.9,4.2);
\node at (8,4.5) {{\large $\mathbf{n}_f$}};
\draw[black,fill](5.6,3.4) circle [radius=0.15];

\draw[blue] (5,0) -- (9,-0.5) -- (9,4) -- (3,5);
\draw[blue] (9,-0.5) -- (12,1) -- (12,5.5) -- (9,4);
\draw[blue] (12,5.5) -- (6,6.5);
\draw[blue, dashed] (12,1) -- (7.5,1.5);
\draw[blue, line width=0.45mm] (7.8,2.65) -- (7.8,2.95);
\draw[blue, line width=0.45mm] (7.65,2.8) -- (7.95,2.8);
\node[blue] at (8.3,2.8) {{\large $E$}};

\draw[red, -{Stealth[length=2mm]}] (3,3) -- (7.8,2.8);
\node[red] at (6.5,2.2) {{\large $\mathbf{d}$}};
\draw[red,fill](5.4,2.9) circle [radius=0.15];

\draw[green] (0,0) -- (-5,0) -- (-5,4) -- (0,5);
\draw[green] (-5,4)-- (-2.5,5.5) -- (1.75,6.2);
\draw[green, dashed] (-5,0) -- (-2.5,1.5) -- (2,2);
\draw[green, dashed] (-2.5,5.5) -- (-2.5,1.5);
\draw[green, line width=0.45mm] (-1.75,2.7)--(-1.45,2.7);
\draw[green, line width=0.45mm] (-1.6,2.55)--(-1.6,2.85);
\node at (-2.3,2.8) {{\large $W$}};

\end{tikzpicture}
\caption{Schematic representation of two computational cells. In this picture, the cell on the right (in blue) represents the upwind cell.}
\label{fig:cartoon}
\end{figure}
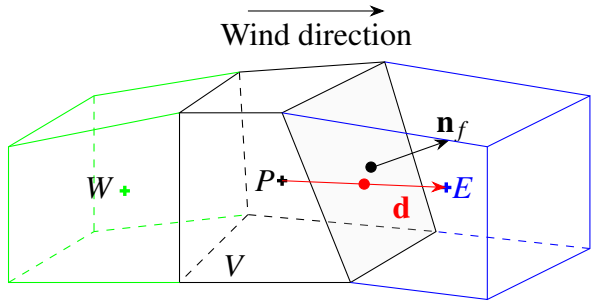


\subsection{Meshing}
Aside from the issue concerning the generation of the geometry, the inherent randomness which characterises most porous systems also affects the generation of the computational mesh.
Far from being a marginal part of the workflow, a fundamental trade-off in the setup of the computational model has to be evaluated, weighing a numerically satisfactory mesh with the increasing costs in machine time for a finer discretisation, both for the mesh generation itself and the consequent flow/transport problem.

Studying a satisfactory portion of a porous medium (i.e.: a \textit{representative elementary volume}) means having to deal (in most cases) with a random system, which immediately excludes the possibility of employing a structured (or block-structured) mesh, usually favoured for their efficiency.
Moreover, one often has to deal with grids composed of a number of elements up to tens of millions \citep{icardi2014pore} or much more \citep{mosby2016computational}, making any kind of ``naked-eye'' inspection unfeasible.

As a result, it is essential to develop a robust pipeline for both the generation of the mesh, which can be controlled by an a-priori discretisation strategy (not reliant on qualitative analysis of the resulting grid), and an efficient process of testing for error convergence based on the simulation results.
Here, we have tested two different unstructured mesh generators, both present in the \textsf{OpenFOAM} suite: \textsf{snappyHexMesh} and \textsf{foamyHexMesh} (hence respectively sHM and fHM for brevity). More details on these tools are available in the appendix while a test-case with the results of the meshing algorithms and their effects on the numerical solution are reported in Section~\ref{sec:mesh}.
We note, for the interested reader, that some valid meshing alternatives can be found in \textsf{Cubit} or \textsf{Gmsh}~\cite{gmsh}, the latter being a freely available and consolidated tool.

%
\section{Results}
\label{sec:results}
\subsection{Transport in periodic sphere packings}\label{sec:BCC}

One of the fields which most commonly deals with the study of transport in porous media is filtration theory. For example, this has been applied for the investigation of aquifer contamination and remediation, \citep{Krol2013,Rolle1,messina2015,Crevacore2016271}.
Correspondingly, a lot of effort has been expended in the past decades to approach this problem, both experimentally and via theoretical investigation.
These kind of problems are basically equivalent, in their simplest formulation, to linear advection-diffusion equations.
Upscaling techniques are therefore robust and well-known~\citep{hornung2012homogenization,bearBook}. However, despite the simplicity of the problem, some issues are still unsolved, such as the quantitative understanding of the role of geometrical parameters in transport processes or the effect of non-linearities.
In recent years, computational studies have begun to complement the available analytical and empirical results, especially when the physical problem presented further difficulties (e.g.: mixing controlled reactions, heterogeneous materials, etc.) which rendered an a-priori upscaling process exceedingly difficult, if not outright impossible~\citep{battiato2011applicability}.

As mentioned elsewhere, the pre-processing step constitutes a big part of a CFD case of transport in porous media.
While in the next sections we explored the process of the generation of the geometrical model, here we will explore the issue of the choice of the optimal meshing strategy and its consequences in the convergence of the finite volume scheme.
To this end, we will consider here the Stokes flow over a unit cell of a periodic arrangement of spheres.
Then, in Section \ref{sec:lagrang}, we turn to look to the upscaling step, and specifically on how to conveniently solve unsteady transport Eulerian equations and on an alternative Lagrangian approach to extract residence-time distribution curves for a solute transport problem.

\subsubsection{Unstructured meshing generation}\label{sec:mesh}
Tests were performed meshing a body-centred cubic structure, where grains were spheres of equal size \citep{P006}, and the medium porosity was held fixed at $0.366$ (i.e.: close to the minimum porosity, but sufficiently high to avoid issues due to the meshing of contact points between adjacent spheres).
Four different meshing strategy were then employed:
\begin{itemize}
\item \textsf{snappyHexMesh} with uniform cell size (\shmU);
\item \textsf{snappyHexMesh} with one level of refinement (\shmR);
\item \textsf{foamyHexMesh} with uniform cell size (\fhmU);
\item \textsf{foamyHexMesh} with one level of refinement (\fhmR).
\end{itemize}

\begin{figure}
    \centering
    \includegraphics[width=.48\textwidth]{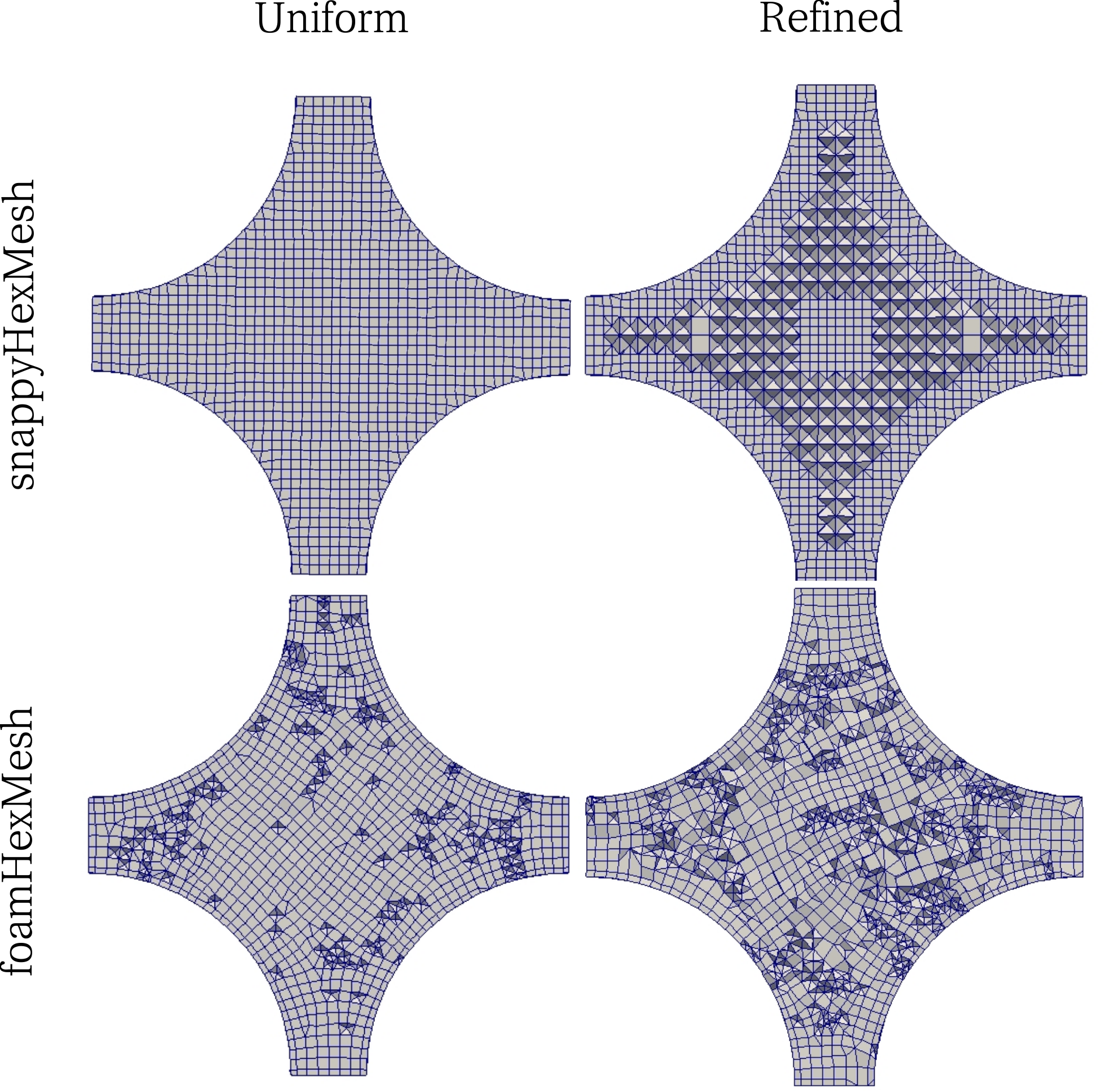}
    \caption{Comparison of four meshing strategy: sHM\_U (up-left), sHM\_R (up-right), fHM\_U (low-left), fHM\_R (low-right). Each mesh appearing in this picture has a total number of cells that is a little less than $5\times10^5$.}
    \label{fig:meshes}
\end{figure}

A representation of these different meshing strategies can be found in Fig.~\ref{fig:meshes}.
When present, the refinement level is located around the grain surface, in order to allow a better discretisation of the boundary layers (of the momentum and concentration/temperature gradients).
For each meshing strategy, meshes of different cell size were built, paying attention to both the mean cells size and the total number of cells.
This latter parameter, see Table \ref{tab:1}, was used to analyse the grid convergence results.
Technical details on the meshing generation are reported in the Appendix.
Equations (\ref{eq:Navier-Stokes}) and (\ref{eq:mass-balance}) were solved imposing a pressure drop between two opposite faces of the domain (along the $x$-axis). The resulting (mean) velocity is thus comparable with environmental applications (Re $\approx$ 3).

\begin{table}
\caption{Total number of cells for ten increasingly fine meshes. The four columns report cell number for an uniform and a wall-refined meshing strategy for \textsf{snappyHexMesh} and \textsf{foamyHexMesh} respectively.}
\label{tab:1}      
\begin{tabular}{lrrrr}
\hline\noalign{\smallskip}
  & \shmU & \shmR & \fhmU & \fhmR \\
\noalign{\smallskip}\hline\noalign{\smallskip}
1  &  1296   &   1296   &   1278   &   914 \\
2  &  3744   &     -    &   3768   &   2799 \\
3  &  5952   &   4176   &   5991   &   4944 \\
4  &  11376   &   8016   &   11509   &   10133 \\
5  &  22800   &   21232   &   22792   &   22919 \\
6  &  44976   &   48288   &   44987   &   46638 \\
7  &  86352   &   84280   &   87856   &   90763 \\
8  &  172944   &   168688   &   182910   &   169449 \\
9  &  302880   &   309360   &   333254   &   280978 \\
10  &  601352   &   606760   &   698629   &   501542 \\
\noalign{\smallskip}\hline
\end{tabular}
\end{table}

\subsubsection{Fluid flow convergence study}
Using the meshes described above, we solve the Stokes~\footnote{We remind the reader that although in the OpenFOAM implementation the equation actually solved is Navier-Stokes, given that \hl{the simulations are performed in laminar flow, this is equivalent to the simple Stokes problem.}}  problem Eqs.(\ref{eq:Navier-Stokes}) and (\ref{eq:mass-balance}), and analyse the numerical performance of the finite-volume schemes. Technical details on the meshing generation are reported in the Appendix.
\begin{figure}
\includegraphics[width=.48\textwidth]{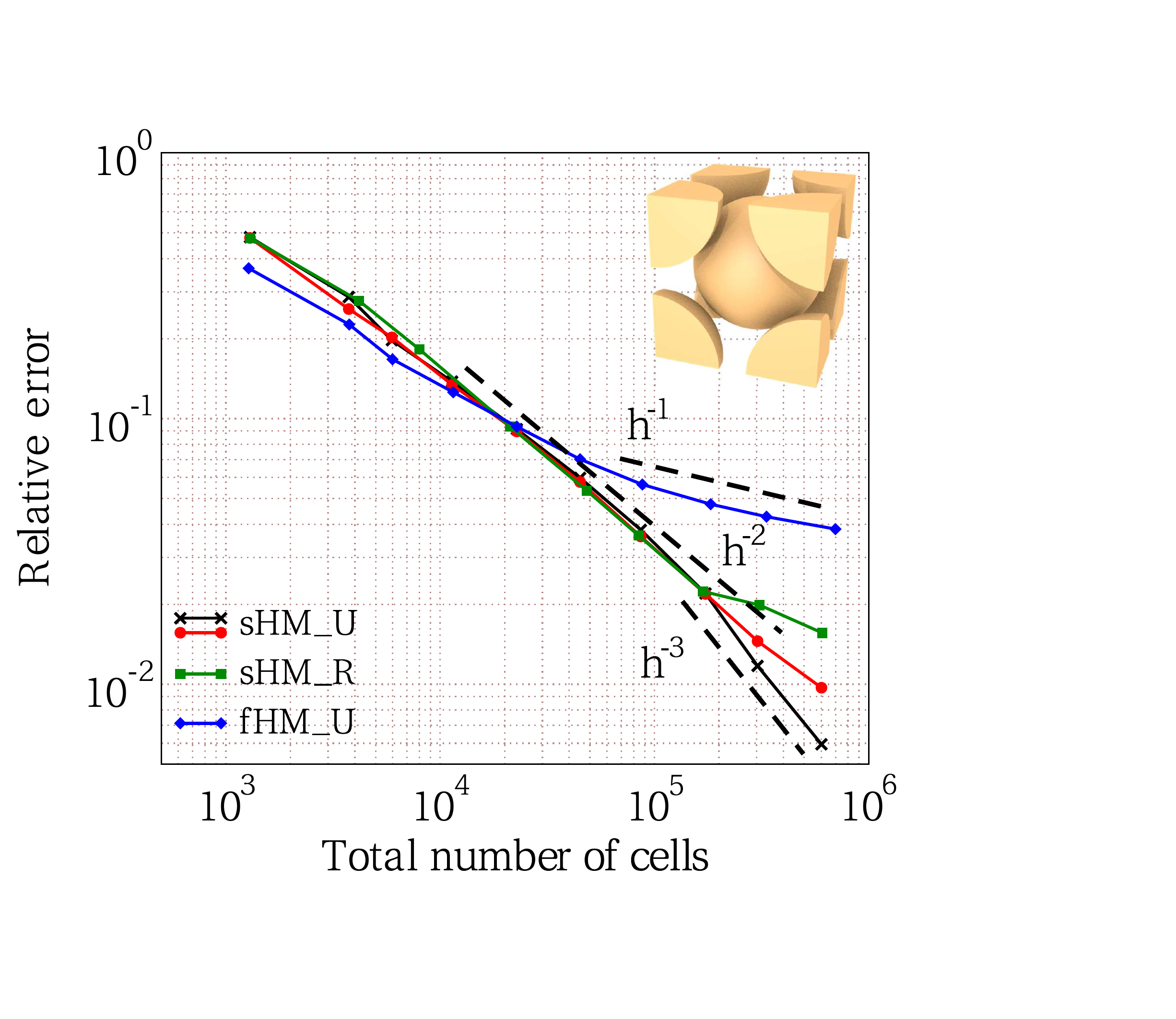}
\caption{Numerical test to verify the convergence properties of the solver for Stokes flow around the body-centred cubic packing (shown in the right upper corner). Relative error in the mean velocity with respect to the total degrees of freedom of the mesh:
uniform mesh with third order schemes (black~$\times$), uniform mesh with second order scheme (red~$\circ$), refined mesh (green~$\Box$), unstructured Voronoi mesh (blue~$\Diamond$). Power law curves are reported according to an equivalent cell size $h$.}
\label{fig:convergence}
\end{figure}
In \fig{convergence} we can observe the effect of the various meshing strategies, combined with different schemes.
The mean velocity (or equivalently the drag coefficient) is compared against a finer resolution result (and reference results in literature \cite{khirevich2015coarse}). 
First of all, it is interesting to notice how in the first part of the plot (i.e.: coarse grids) the main error is always approximately second-order.
This is due to the fact that, in our meshing strategy, the surface of the sphere is not exactly preserved but it is itself discretised.
This can be easily estimated from simple geometrical arguments and the theory of elliptic operators.
This ``geometrical error'' can instead behave very differently in presence of significant boundary-layer effects (Navier-Stokes or boundary reactions).
After this first regime, we can observe that, in general, the uniform meshes generated with \textsf{snappyHexMesh} perform significantly better reaching a second- and third-order convergence, according to the finite-volume scheme chosen.
On the other hand, the refined grid sHM\_R loses the second-order accuracy due to the introduced non-orthogonality and skewness of the cells.
As explained earlier, this can be corrected only with explicit terms that could cause unboundedness.
Therefore we limit the corrections, and obtain and sub-optimal convergence rate.

For this linear problem, the advantage of refining near the boundary does not pay off, as the computational savings are not enough to justify this deterioration of the numerical convergence.
Similarly, the unstructured meshing strategies based on Voronoi tessellation (fHM both uniform and refined), despite being attractive for their robustness and simplicity, are characterised only by a linear decay of the error; please note that fHM\_R is not reported in \fig{convergence} for sake of simplicity, as it practically behaves like the fHM\_U. However they seem to perform better for very coarse meshes where the geometrical features can be better represented with a small number of Voronoi tiles.

\subsubsection{Residence time distribution and breakthrough curves}\label{sec:lagrang}
The solute transport problem can be solved, either in a Lagrangian or Eulerian way, given the computation of the flow field (from now on, we will refer only to the results obtained with the finest uniform mesh with second-order schemes). 
For infinite P\'eclet numbers (purely hyperbolic equation), the transport can be fully characterised by the streamlines.
This is particularly convenient as, within an Eulerian formulation, specific advection schemes and highly refined meshes would be needed to avoid significant numerical artificial diffusion or instabilities~\cite{benson2017comparison}.
Streamlines are integrated with the \textsf{streamLine} post-processing tool available in OpenFOAM 4, and they are generated seeding a cloud of points that can be located wherever within the domain.
In particular, we provided points seeded on the inlet face of the domain, perpendicular to the flow direction.
Points can be placed either randomly or on a regular grid; the former strategy should be preferred to provide statistically reliable samples.

In presence of finite (possibly very small) diffusion, a linear drift diffusion Ito's Stochastic Differential Equation should instead be solved.
This requires more sophisticated and expensive random walk algorithms.
These are usually coupled to the fluid flow equations, allowing a possible coupling between the solute concentration and the fluid.
Despite the higher accuracy of Lagrangian methods for advection-dominated transport, the Lagrangian discretisation of more complex non-linear advection diffusion reaction and the coupling with the fluid flow can be non trivial and give rise to significant statistical errors due to the finite number of trajectories.

Here we follow 1000 particle trajectories from the inlet to the outlet and compute the cumulative distribution of arrival (or residence) times $f(t)$ in the pure advection regime, and compare them with Eulerian simulations at increasing P\'eclet numbers.
The cumulative arrival times of a random ensemble of particles initially positioned at the inlet is equivalent to the time integral of the concentration of particles at the outlet with inlet concentration delta-distributed in time (the ``response'' function of an impulse at time $t=0$).
This is therefore equivalent to the time-derivative of the so-called ``breakthrough curve'', i.e. the response to an Heaviside function.

Having established a connection between Eulerian and Lagrangian simulations in the non-reacting case, what we want to study is the appearance of ``anomalous transport'' characteristics when the diffusion is small enough not to allow for a significant mixing, therefore resulting in particles following preferential directions and staying for long times in the same streamline.
This, apart from very special cases (e.g. Gaussian distributed flow field), generates non-Fickian transport behaviour.
This is a well-known problem in porous media who has received significant attention (see, for example, \cite{Dentz:Carrera:2007} and \cite{DentzGouzeAWR2012}).
It can be easily understood if one tries to fit the three-dimensional results with an equivalent one-dimensional linear advection diffusion equation (with constant velocity, equal to the mean velocity, and unknown diffusion\footnote{In this case, it would be more correct to talk about hydro-dynamical \textit{dispersion} as the effective diffusion is caused both by molecular diffusion and by mechanical dispersion due to the heterogeneous velocity field.}).

\begin{figure}
\includegraphics[width=.5\textwidth]{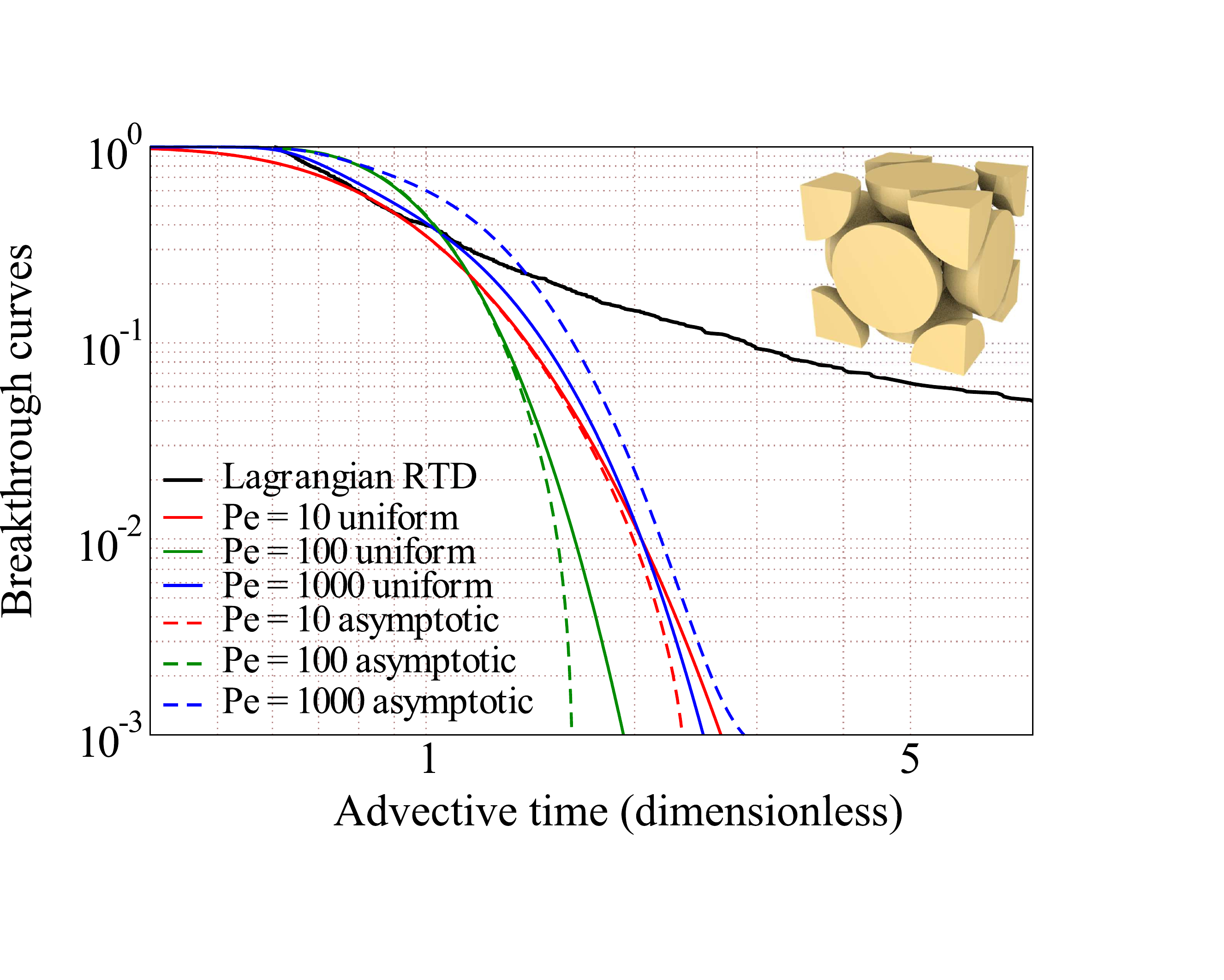}
\caption{Breakthrough curves $f(t)=\spatavg{c}_{u_{x}}$ for a single face-centered cubic cell computed from Eulerian simulations at finite P\'eclet numbers (respectively in red, green, and blue for Pe= 10, 100, and 1000) compared with the Lagrangian residence time distribution (in black). For the Eulerian cases, the continuous lines result from employing the classic boundary conditions, while the dashed lines are the results employing the scalar quasi-periodic boundary conditions described in Section~\ref{sec:bc}. }
\label{fig:RTD}
\end{figure}

For the purpose of this section, simulation were performed on a face-centred body structure (made of spherical grains of equal size) with porosity 0.40.  
In \fig{RTD} the residual concentration $1-f(t)$ is shown for Lagrangian streamlines (equivalent to infinite P\'eclet number) and for P\'eclet numbers ranging from 10 to 1000, against the dimensionless advective time $t/\tau$, where $\tau$ is the arrival time for the advective front (i.e., considering the mean velocity).
At first, a uniform (across the inlet plane) boundary condition has been tested (ref. \fig{RTD} continuous lines).
Here it can be seen that for $Pe=10$ the transport is still dominated by diffusion, with a first arrival time lower than the front arrival time.
Then, for higher P\'eclet numbers, the curves start to show a significant power-law ``tailing'',  while the  solution of a 1D advection diffusion equation would predict an exponential decay.
This is due to the incomplete mixing and preferential trajectories. 
Although it is not in the objectives of the paper to analyse the anomalous transport characteristics, it is interesting to notice here that a significantly different behaviour is observed when the technique and boundary conditions described in \ref{sec:bc} are used (dashed lines).
In fact, when the non reactive stationary quasi-periodic problem is solved within the porous space, an ``asymptotic'' self-similar profile is obtained.
When using this asymptotic profile as the boundary condition at the inlet, the resulting transport becomes clearly ``Fickian'', with exponential tails.
This is because the profile found by the quasi-periodic ``closure'' represents the profile for which the stationary outlet concentration is uniform.
Therefore, it can be considered as an inverse problem of finding the profile that cancels out the preferential trajectories.
The black solid line finally represents the arrival (or transit or residence) times of purely advective Lagrangian trajectories.
As it can be seen, the first arrival times match well with the high P\'eclet case but the absence of diffusion cause a very significant tailing with a power-law behaviour.

\subsection{Flow over porous surfaces}\label{sec:pipeSand}
In this section, we will explore the case of flow and heat transfer in a pipe containing, for a small portion of its volume, a number of spherical solid grains settled on its lower side.
The problem of studying heat transfer in pipes, and consequently evaluating the heat transfer coefficient in pipes has been extensively studied~\citep{Bird1960,incropera2007fundamentals,choi1997momentum,manes2009turbulence,mossner2015}; in this work, we want to show how the presence of randomly arranged solid matter affects fluid flow and heat transfer.
We compare the results with the analogous ``clean'' cases where no deposited matter is present.
In particular, this section will serve as a proof-of-concept for the methodology we employed for both the generation of the geometry comprising the settled matter (as described in the following paragraphs) and the easiness with which it is possible to explore geometrical parameter space and operating conditions.
Then, an overview of the numerical details of the performed simulations, as well as the results obtained, are presented.

\subsubsection{Geometry generation}

As it has been mentioned, all the geometrical models used in this work were built \textit{in-silico}, avoiding the costly step of obtaining experimental samples and adapting them for use in CFD codes.
Specifically, the case study presented in this section was created with \textsf{Blender}, a free and open-source software used in computer graphics applications, which can be used to perform rigid-body simulations.

The settled matter is represented by uniformly-sized spheres, which were chosen to be of a diameter equal to one-twentieth of the pipe diameter. 
In the \textsf{Blender} simulation setup, grains are first located above the container (which is now represented by just the bottom half of the pipe, to let the granular matter settle into it), taking care of adding randomness to their initial position in order to avoid unwanted ``arranged'' structures in the final packing.
Then, the solid grains are free to fall due to the effect of gravity, which is the only outside force. The final state of the system is given by the solution of the rigid-body problem, where the solid grains are treated as non-deformable and impenetrable entities\footnote{More precisely, the definition for the \textit{rigid bodies} is given as solids possessing an infinite repulsive potential present on their surface.}, and the only interactions considered are instantaneous collisions and sliding.
After reaching the final state, identified by the absence of relative motion between all the solid grains and the pipe, the geometric model is exported for its use in the CFD code (substituting the half-pipe with the complete pipe for the simulations).
This kind of methodology has been successfully used to represent both packings of catalyst particles in chemical reactors~\citep{P003,boccardo2019fine} and in digital rock physics applications~\citep{icardi2016predictivity}: the numerical details of the rigid-body simulations can be found therein.

With the process just described, we created three different geometries, each characterised by the (increasing) volume of settled matter, roughly corresponding to an average height, $H_g$, of the settled matter packings equal to two, three, and four grain diameters respectively. 
In all cases, the pipe length is equal to three times the pipe diameter.
The spatial domain was discretised with a sHM\_R strategy.
Also, an in-depth description of the use of this mesh generator in the case of random packings and an analysis of mesh convergence for the study of advection-transport problems can be found in our earlier work~\citep{IBMTS2014}.


\subsubsection{Numerical details}

\begin{figure}[h!]
\includegraphics[width=.5\textwidth]{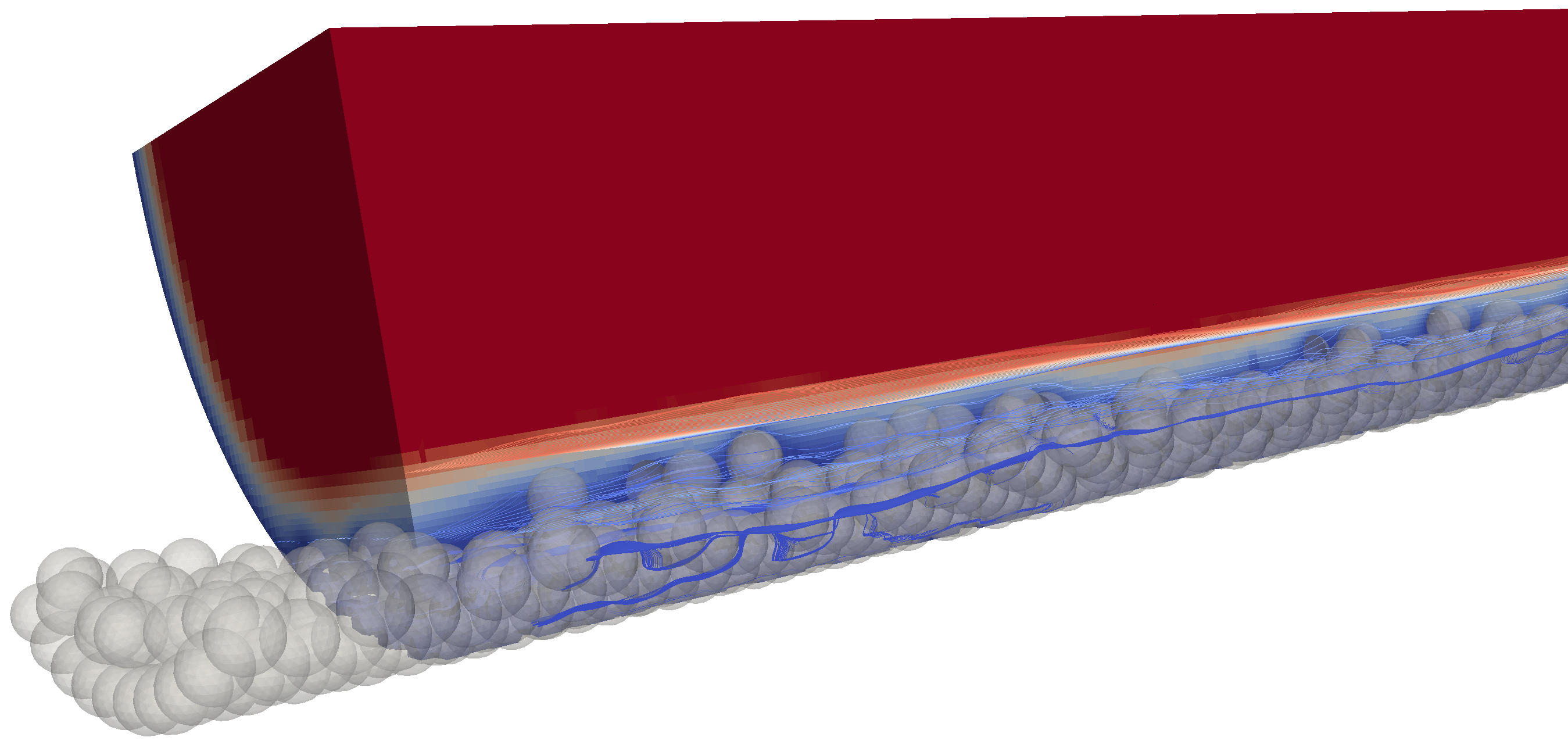}
\caption{Velocity contour plot inside the pipe and fluid streamlines (coloured by fluid velocity) inside the granular portion of the domain, showing the increasing distortion of the fluid paths moving from the bulk towards the bottom of the pipe.}
\label{fig:pS}
\end{figure}

The first step, as for the case described in the previous section, has been to solve the equation for the motion of fluid inside the pipe.
We thus solved Eqs.(\ref{eq:Navier-Stokes}) and (\ref{eq:mass-balance}) in laminar regime by setting a constant pressure drop between the inlet and the outlet of the domain, corresponding to a Reynolds number Re $\approx$ 1 (we used again second-order schemes).
A snapshot of the velocity contour plot inside the pipe can be found in Fig.~\ref{fig:pS}, together with a representation of the slower fluid streamlines inside the  volume occupied by the settled granular matter.

\begin{figure}[h!]
\includegraphics[width=.5\textwidth]{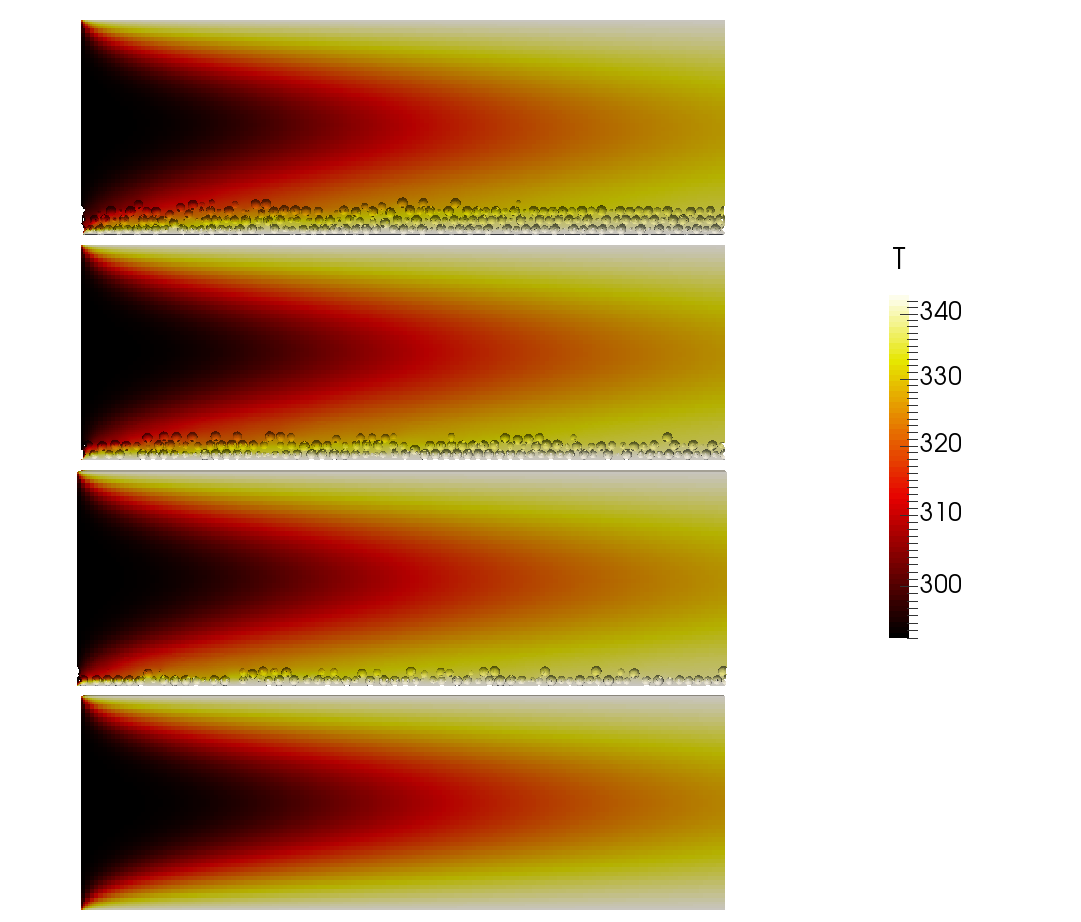}
\caption{Temperature contour plots on an axial plane of the pipe for cases with average  granular matter height from the bottom of the pipe equal to roughly four, three, and two grain diameters (from top to bottom), and for the ``clean'' pipe (bottom).}
\label{fig:pS-contourT}
\end{figure}

A simulation of heat transfer was then performed solving Eq.~\eqref{eq:heatTransfer}, setting a constant fluid temperature at inlet, equal to T=293 K , and a zero flux at the outlet.
A Dirichlet condition of constant temperature T=343 K was set on the pipe wall, while a zero thermal flux (corresponding to $\nabla T=0$) was set on the grains.
This choice of boundary conditions is equivalent to assume a situation of perfect thermal equilibrium between the flowing fluid and the solid grains: this is justified in the limiting cases of infinitely fast heat transport in the solid, if one considers the long-time, steady-state, solution of the problem.
A snapshot of the contour plots of fluid temperature on the median length-wise section of the pipe\footnote{More precisely, taken on a plane parallel to both the main flow direction and the direction of gravity, positioned on the pipe axis.} is found in Fig.~\ref{fig:pS-contourT}, which shows the amount of solid matter in each case and the slight difference in the thermal boundary layer at the bottom.

In the next section, we will instead explore the problem of pure heat transfer in solids, between a continuous matrix and a disperse solid phase, characterised by two different thermal diffusivity values: these two cases are meant to represent, in a way, two opposite situations in the study of heat transfer in heterogeneous materials.

\subsubsection{Heat transfer coefficient}

When the steady-state solution of the problem of advection-diffusion has been obtained, the heat fluxes along the main flow direction are extracted in order to calculate the local heat transfer coefficient $h_{loc}$ \citep{Bird1960} evaluated at a number of successive planes orthogonal to the main flow direction (i.e. the pipe axis):

\begin{equation}\label{eq:h_loc}
h_{loc}=\dfrac{q \rho C_p}{\pi D}\dfrac{d T_{bulk}(x)}{d x}\dfrac{1}{T_{wall}-T_{bulk}}
\end{equation}

where $q$ (m s$^{-1}$) is the average flow velocity, $\rho$ (kg m$^{-3}$) its density, $C_p$ (J kg$^{-1}$ K$^{-1}$) its specific heat capacity, $x$ (m) the distance of the averaging plane from the inlet boundary, and $T_{wall}$ and $T_{bulk}$ (K) respectively the temperature on the wall of the pipe and the local average of the fluid bulk temperature.

\begin{figure}[h!]
\includegraphics[width=.5\textwidth]{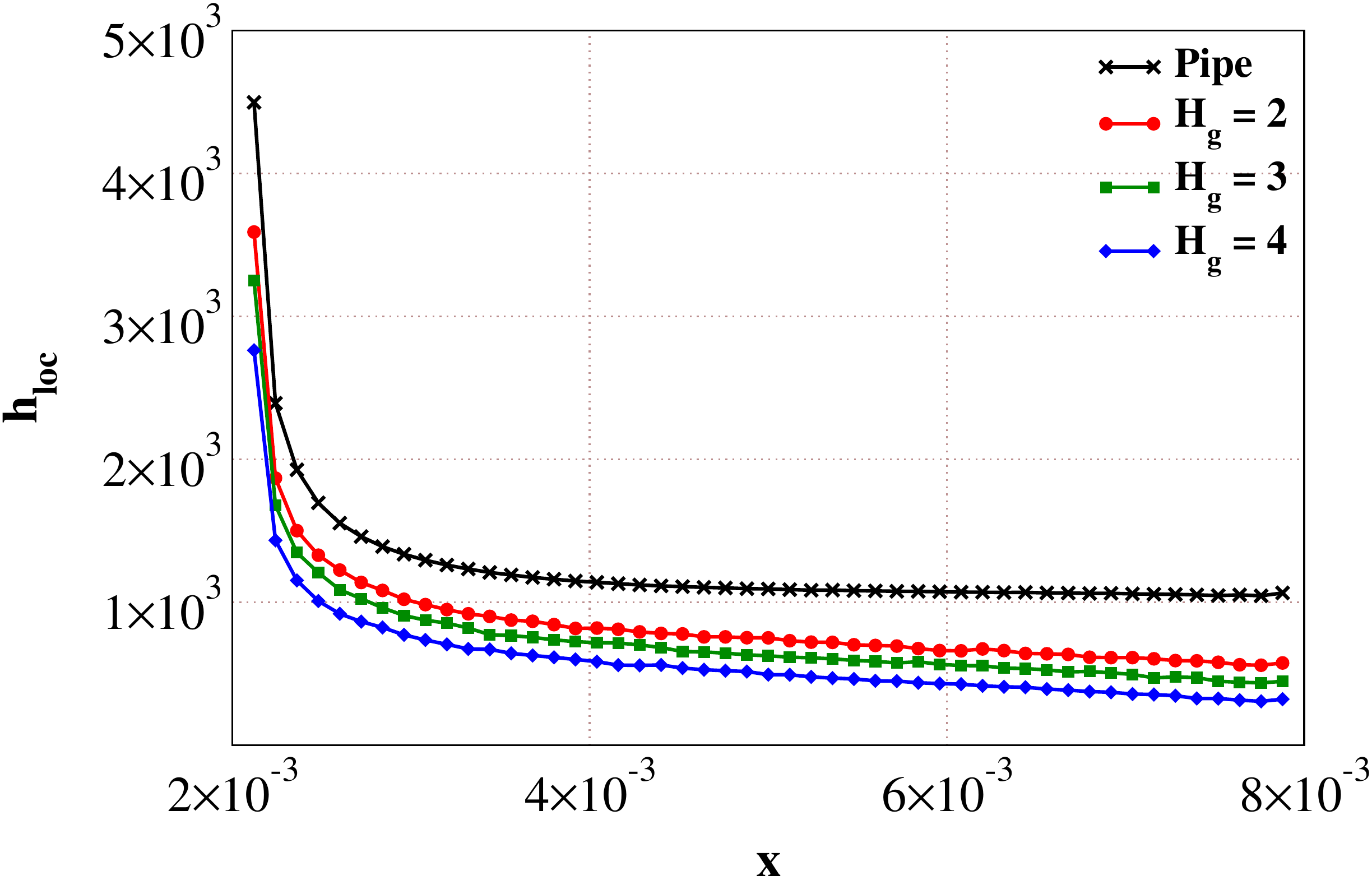}
\caption{Heat transfer coefficient $h_{loc}$ as a function of the distance from the inlet (in meters): values for the ``clean'' pipe (black~$\times$, line at the top) and for granular matter height from the bottom of the pipe equal to roughly two, three, and four grain diameters (respectively red~$\circ$, green~$\Box$, and blue~$\Diamond$ curves, in descending order in the graph).}
\label{fig:h_loc}
\end{figure}

The $h_{loc}$ results for each case are shown in Fig.~\ref{fig:h_loc}, compared with the heat transfer coefficient of the ``clean'' pipe devoid of settled matter.
As it can be seen, the presence of granular solids at the bottom of the pipe cause two effects.
The first is that of reducing the total heat transfer coefficient (and consequently Nusselt number), which is smaller the bigger the volume of settled particles: this is due to the reduction of the contribution of advective heat transport with respect to conductive transport, caused by the much lower fluid velocities in the inter-granular portion of the domain.
Then, it can also be noticed that the presence of randomly arranged grains have an effect on the dynamics of reaching an asymptotic regime: while in the ``clean'' pipe case the heat transfer coefficient reaches a plateau not far from the inlet (as it is expected), for the cases with granular matter it is clear that a decreasing trend is still present even close to the domain outlet.

\subsection{Heat transfer in heterogeneous materials}
\label{sec:soliDisp}
In the preceding section, we only considered the effect of the presence of a granular solid on flow and heat transfer in pipes, neglecting the heat transfer between the bulk of the fluid (or the pipe wall) and the grains themselves.
In this section, we will instead explore a case where no fluid flow is present, investigating the case of heat transfer between two solid phases with different thermal diffusivities $\alpha_{1}$ and $\alpha_{2}$, the first phase constituting a continuous matrix and the second phase dispersed (as solid inclusions) in the first; moreover, the diffusivity in the continuous matrix is anisotropic, having different values for its longitudinal and transversal components.

\subsubsection{Extended Jodrey-Tory algorithm (EJT)}
For this application, a periodic extension of the classical Jodrey-Tory algorithm \cite{jodrey1981computer} has been implemented to deal with randomly oriented ellipsoids.
Despite not being computationally optimal for low-porosity (high-density) materials, it can easily generate random packings with a specific porosity and a specific overlapping between grains, as well as clustering.
The new algorithm has been implemented in the open-source repository \textsf{mlmc-porescale}\citep{icardi2016predictivity}, and can be sketched as follows:
\begin{enumerate}
\item The user specifies a domain size (including periodicity information), a target porosity, the minimum degree of overlapping $\theta$ ($\theta<1$ allows overlapping, $\theta>1$ prescribe a certain distance from grain to grain), a maximum distance between grains $\Theta>\theta$ (this is enforced on clusters of $N_{\Theta}$ particles), and a maximum limit of particle numbers and displacement iterations.
Furthermore, geometrical statistics about the ellipticity of the grains and their size have be to provided.
In the simplest (isotropic) case, a single random variable is selected (log-normal, truncated Gaussian or uniform) to describe the ellipses canonical axes length.
\item Random ellipsoids are generated until they reach the desired porosity (not considering overlapping).
For each ellipsoid, a random unitary matrix $C$ is sampled~\citep{ozols2009generate} according to the axes lengths and orientation statistics.
This is then decomposed $C=C_{1}C_{2}$ into a rigid rotation part $C_{1}\in SO(3)$ and a diagonal part $C_{2}$(responsible for scaling).
\item Once all ellipsoids have been randomly placed in the box, a greedy-type algorithm is executed: at each iteration, the pair-wise distances are generated and the first $N_{moves}$ pairs, characterised by the largest overlapping ratio, are detached along the line of the pair-wise distance axis.
The displacement length controls the convergence properties of the algorithm but its optimal value strongly depends on the porosity of the packing.
Usually a relative displacement of $0.2-0.5$ times the particle size is chosen.
\item If a maximum distance is specified, the same type of moves are applied to shorten the pair-wise distance.
Here, however, this can be only applied to a specific cluster of particles selected as the closest ones.
\item The moves are applied iteratively until all the pair-wise distances satisfy the criteria.
Periodicity is ensured by implementing the appropriate toric metric to compute the distances.

\end{enumerate}

\hl{It has to be noted that there is not ensured stopping point for the algorithm and a total number of moves of the greedy algorithm described in point 3. is fixed as an upper bound.
For a given available volume and number of ellipsoids, in the current implementation of the code, one needs to choose the overlapping layer appropriately and large enough for a packing solution to
exist.
In the future, however, simple adaptivity could be implemented to override the settings for overlaps or for ellipsoid size, to converge faster to a final arrangement.}

\begin{figure}[h!]
\includegraphics[width=.5\textwidth]{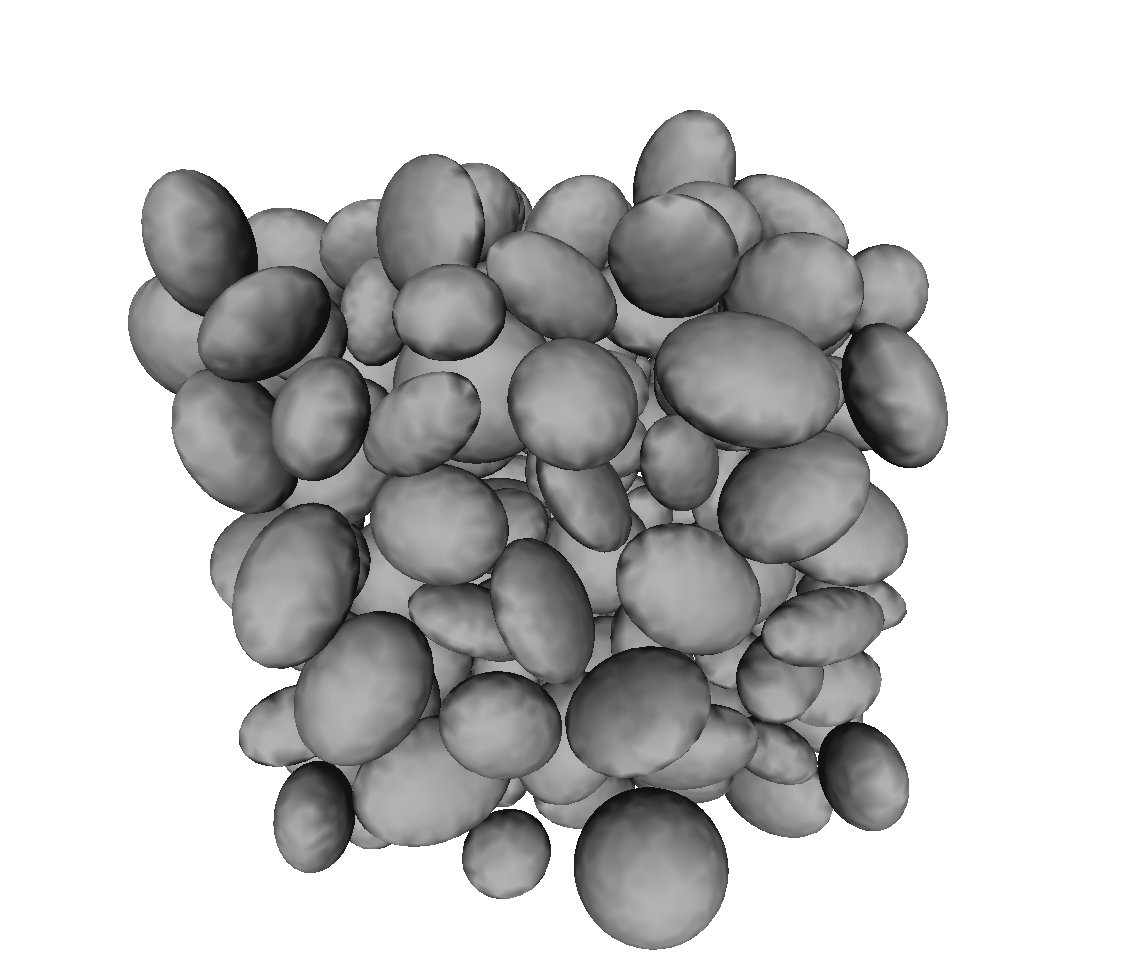}
\caption{Example of random ellipsoids with random (statistically uniform in $SO(3)$) orientation, generated with the Extended Jodrey-Tory algorithm (EJT).}
\label{fig:eJT}
\end{figure}

An example of 200 touching ($\theta=1,\,\Theta=\infty$) ellipsoids in a periodic arrangement with 0.45 porosity is shown in \fig{eJT}.

\subsubsection{Effective diffusion}
The micro-scale heat transfer problem to be solved is \eq{heatTransfer} with $\mathbf{u}=0$ and
$$
\alpha(\xb)=
\begin{cases}
\mbox{diag}(0.1, 10, 10) \quad \mbox{for}\; \xb\in\Omega_{1}\;,\\
\mbox{diag}(1, 1, 1) \quad \mbox{for}\; \xb\in\Omega_{2} 
\end{cases}
$$
Where diag is the diagonal matrix constructed from the vector. This represents the case of a continuous phase made of highly conductive layers (e.g., graphene, therefore with large conductivity in only two directions), reinforced with ellipsoidal inclusions with isotropic conductive properties.

\begin{figure}[h!]
\includegraphics[width=.5\textwidth]{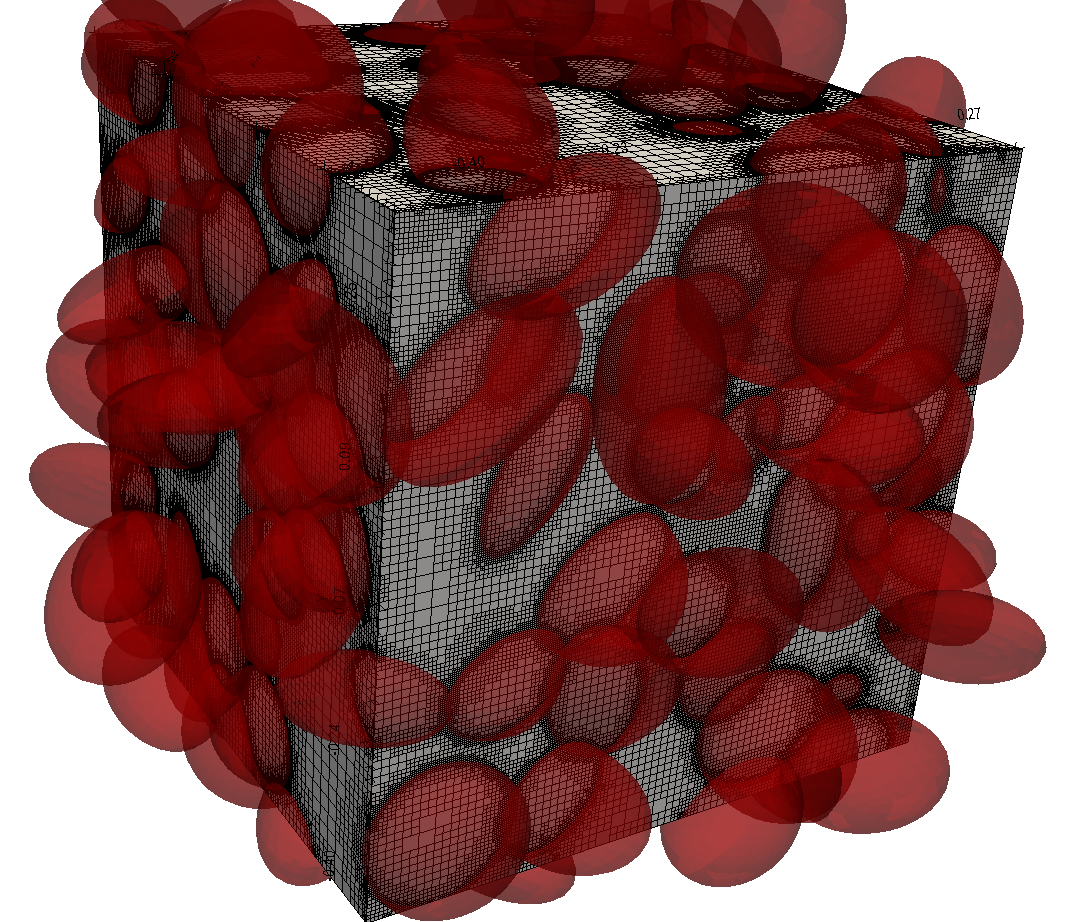}
\caption{Representation of the complete geometry, showing the mesh of two distinct domains, with the dispersed solid inclusions generated with the Extended Jodrey-Tory algorithm highlighted as transparent red volumes.}
\label{fig:eJT-mesh2}
\end{figure}

In \fig{eJT-mesh2}, the periodic structure (with the same properties of the one in \fig{eJT} but with up to a 20\% overlapping allowed) is shown with the underlying locally refined mesh (\shmR).
We have shown above that the local refinements destroys the second-order convergence of the numerical schemes.
However, due to the discontinuous diffusion coefficient, a second-order convergence would not be possible with our numerical schemes.

\begin{figure}[h!]
\includegraphics[width=.5\textwidth]{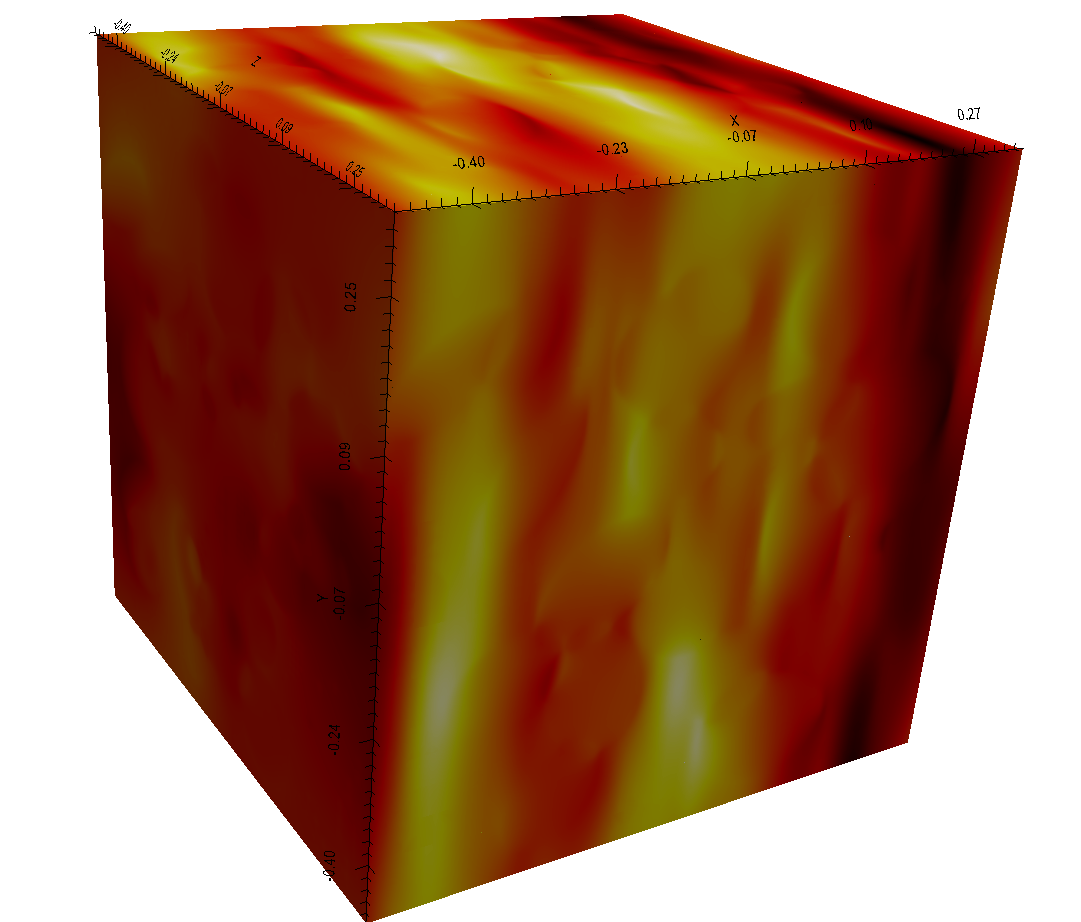}
\caption{Local temperature fluctuations $c-\mathbf{p}\cdot\xb$ for the the quasi-periodic cell problem. The volume average of their gradient gives the effective diffusion coefficient.}
\label{fig:eJT-mesh3}
\end{figure}

To compute the effective (homogenised) diffusion coefficient, one may follow a volume averaging strategy, applying the quasi-periodic BCs in \eq{cons-bc} (results shown in Fig.~(\ref{fig:eJT-mesh3})).
This is more convenient than solving the homogenisation cell problem where the derivative of the discontinuous conductivity appears as a source term of the equation~\citep{hornung2012homogenization}.
The resulting effective diffusion is, in this case:
$$
\alpha_{eff}=\matrix{4.955 & \epsilon & \epsilon\\\epsilon & 39.86 & \epsilon\\ \epsilon & \epsilon & 38.41}
$$
where $\epsilon$ represents small coefficients $\approx10^{-3}$. As it can be seen, the resulting effective diffusion, as expected, is again anisotropic.

\section{Conclusions}
In this work, we have shown the computational challenges involved in the simulation of transport problems in the field of porous media or, more generally, in problems which feature multiple \textit{spatial} scales, or whose evolution takes place over multiple \textit{temporal} scales: examples of which we have respectively provided in the form of the case of a pipe with small-size granular sedimentation, and the study of solute transport at very high P\'eclet numbers.
Moreover, we have also presented applications of a number of third-party open-source codes and showcased our freely available extensions. These range from tools used to quickly and robustly generate a wide variety of three-dimensional realistic porous media models, to  innovative computational setups providing for a great decrease of the computational expense in quasi-periodic cases~\cite{P006}.
In the first case, for example, an extended Jodrey-Tory algorithm has been developed, to generate random packings of ellipsoids, while, in the latter, a simple steady-state closure problem, fully consistent with volume averaging and homogenisation, is derived for time-dependent transport with linear (and constant in time) transport parameters.
Then, the semi-periodic setting allows to solve one single periodic cell, while a Representative Elementary Volume is, in the case of advection-diffusion, dependent on the P\'eclet numbers of the system, and possibly including tens or hundreds of periodic cells.

Given our focus on free and open-source software and having primarily chosen, due to its robustness and flexibility, the \textsf{OpenFOAM} platform to discretise our models, we have illustrated the main features of the meshing and the discretisation schemes underlying its classical finite-volume approach.
We have explored the alternatives provided in the \textsf{OpenFOAM} suite regarding the choice of the meshing pre-processors, providing an in-depth analysis about the order of relative error convergence of the two approaches, while also exploring different meshing strategies.
Again, this is done with the objective of providing guidance and an initial benchmark with respect to the pre-processing part of the numerical simulation, which probably has the greatest impact on the accuracy of the final results of the model.

The simulations presented, although representing three-dimensional, realistic, and physically meaningful systems, do not represent real large-scale structures and therefore do not require any sophisticated adaptivity and parallelism.


In this work therefore, we have chosen to limit our investigation to showing the difference between two radically different body-fitted approaches, a Cartesian-based cut-cell approach and a Voronoi unstructured approach, presenting the differences in accuracy between the two, \hl{with the focus on building a general-purpose package for medium-scale simulation.}
\textsf{OpenFOAM} has, however, a very flexible parallel structure and, as such, most of our solvers are inherently parallel, with some extra functionality (particularly for I/O and averaging operations) implemented in a parallel way. Depending on the overall size of the geometry, this can be easily be parallelisable, with good scaling properties, up to a few hundreds cores.
\hl{It has to be noted that, while OpenFOAM can easily be parallelised on many thousands cores (although, depending on the specific solver, losing the optimal scaling), in our experience, a multi-scale micro-macro offline coupling could be a more viable and robust option to deal with porous media.}

Finally, the aggregate results of this work demonstrate how it is possible to build a completely computational pipeline for the study of a variety of physical problems, from the generation of a suitable and realistic geometric model, to the actual simulation run, to the extraction of the relevant data. This already found many applications in the field of environmental and chemical engineering and will be extended for other applications, for example, in energy storage.
This emphasis is meant and presented both as an alternative between free and open-source software as opposed to licensed and ``closed'' software, and between \textit{in-silico} methods of reconstruction of the geometry of the porous medium as opposed to more costly (and often less available) methods such as micro-computed-tomography or x-ray imaging.
In both cases, the clear purpose is to work towards the development of open scientific frameworks providing easily reproducible and verifiable results that links rigorous mathematical models and numerics with physics and engineering applications.

\begin{acknowledgements}
MI acknowledge the financial support of AVL and the computational resources provided by the Warwick Centre for Scientific Computing.
\end{acknowledgements}

\bibliographystyle{spmpsci}      
\bibliography{depBib}   

\appendix

\section{\textsf{OpenFOAM} meshing details}
In this appendix, we explain the features of \textsf{foamyHexMesh} (\fhm) and \textsf{snappyHexMesh} (\shm) used to generated the meshes of all our numerical results. 
\fhm was introduced in the 2.3.0 version; it is based on a Delaunay tessellation, and the resulting Voronoi dual is typically a hex-dominated mesh \cite{oF}. This mesher tool was thought to provide a good feature conformation, avoiding high aspect ratio cells, \cite{oF}.
To the best of our knowledge, detailed guidelines for the usage of \fhm for porous materials are not available. Thus in the following of this section, prior to provide the parameters used to generate the presented meshes, we present a quick description of the key  parameters of \fhm, while making a parallelism with \shm, highlighting advantages and disadvantages of these two meshing tools.
Both \shm and \fhm deals with in-built analytical surfaces and \textsf{.stl} files. Analytical surfaces (spheres, boxes, cylinders) are readily implementable in \shm, while this is not true in \fhm, where the user must explicitly manage the interception of the surfaces with the (domain) bounding box (or among the surfaces themselves when necessary); thus the usage of externally generated geometries is of easier implementation. 
In this work, when needed, \textsf{.stl} geometries have been generated with \textsf{openSCAD} \cite{oS}.
With \fhm the use of \textsf{.stl} allows a straightforward implementation of the surface refinement, as it is sufficient to include an extra \textsf{+.stl} file (or a \textsf{.stl} composed of different regions/surfaces) to address those surface on which a greater refinement (moving from the surface to the bulk region) is required.\\
With \shm the dimension and number of cells, initialised in the starting Cartesian mesh properties, encoded in the dictionary \textsf{blockMeshDict}, can be manipulated with \textsf{refinementRegions}.
In \fhm, the way cell dimensions can be controlled depends instead on the \textsf{initialPointsMethod} selected. In particular, we tested both the \textsf{uniformGrid} and the \textsf{autoDensity} functions. In the former case, the cell size (and thus the total number of cells) is controlled fixing the \textsf{initialCellSize} parameter, while in the \textsf{autoDensity} case it is possible to define \textsf{defaultCellSize}, in the \textsf{motionControl} section\footnote{Notice that when the \textsf{uniformGrid} method is used, \textsf{motinControl} fields are practically ignored, particularly if \textsf{initialCellSize} $\gg$ \textsf{defaultCellSize}.}. In both cases, the size variable makes reference to the mean cell side length (keeping in mind that the mesher try to produce hexaedral cells). Notice that, since \shm relies on a Cartesian approach, while \fhm does not, this implicitly implies that \shm usually allow a better control on the final number of cells per grains diameters. 
Another parameter of interest in \fhm is the \textsf{maxSmoothingIterations} number (\textsf{motionControl} section). It is intuitive that the higher is the number of iterations, the higher is the mesh quality, and this is particularly true when going from single core to parallel run of \fhm, as this last approach requires a greater number of iterations to achieve meshes of comparable quality.\\

The  most important parameters used to generate the meshes in Section \ref{sec:BCC} are reported below:\\
\paragraph{\shmU}:
\begin{itemize}
\item \textsf{blockMeshDict} $\to$ \textsf{blocks}: number of cells.
\item \textsf{snappyHexMeshDict} $\to$ \textsf{refinementSurfaces}: \textsf{level (0 0)}.
\end{itemize}

\paragraph{\shmR}:
\begin{itemize}
\item \textsf{blockMeshDict} $\to$ \textsf{blocks}: number of cells.
\item \textsf{snappyHexMeshDict} $\to$ \textsf{refinementSurfaces}: \textsf{level (2 2)}.
\end{itemize}

\paragraph{\fhmU}
\begin{itemize}
\item \textsf{initialPoints}: \textsf{initialPointsMethod \quad autoDensity}.
\item \textsf{motionControl}: \textsf{defaultCellSize}.
\item \textsf{motionControl}: \textsf{minimumCellSizeCoeff} \quad 10\% \textsf{defaultCellSize} value.
\item \textsf{motionControl}: \textsf{maxSmoothingIterations} 500.
\item \textsf{motionControl}: \textsf{maxSmoothingIterations} 0.
\end{itemize}

\paragraph{\fhmR}
\begin{itemize}
\item \textsf{initialPoints}: \textsf{initialPointsMethod \quad autoDensity}.
\item \textsf{motionControl}: \textsf{defaultCellSize}.
\item \textsf{motionControl}: \textsf{minimumCellSizeCoeff} \quad 10\% \textsf{defaultCellSize} value.
\item \textsf{motionControl}: \textsf{maxSmoothingIterations} 500.
\item \textsf{motionControl}: \textsf{maxSmoothingIterations} 0.
\item \textsf{motionControl} $\to$ \textsf{shapeControlFunctions} $\to$ surfaceToBeRefined $\to$ \textsf{uniformValueCoeffs}: \textsf{surfaceCellSizeCoeff} \quad 0.5;
\end{itemize}

\end{document}